\documentclass[amsmath,amssymb,
 reprint,%
]{revtex4-1}

\usepackage{graphicx}
\usepackage{dcolumn}
\usepackage{bm}
\usepackage{hyperref}
\usepackage[dvipsnames]{xcolor}
\hypersetup{colorlinks=true,citecolor=Blue,linkcolor=black,urlcolor=MidnightBlue}

\usepackage[utf8]{inputenc}
\usepackage[T1]{fontenc}
\usepackage{mathptmx}
\usepackage{xcolor}

\usepackage[title]{appendix}

\usepackage{comment}

\begin{document}

\newcommand{\1}{{\bf \scriptstyle 1}\!\!{1}}
\newcommand{\unit}{\overleftrightarrow{{\bf \scriptstyle 1}\!\!{1}}}
\newcommand{\I}{{\rm i}}
\newcommand{\p}{\partial}
\newcommand{\D}{^{\dagger}}
\newcommand{\hbe}{\hat{\bf e}}
\newcommand{\bfa}{{\bf a}}
\newcommand{\bx}{{\bf x}}
\newcommand{\hbx}{\hat{\bf x}}
\newcommand{\by}{{\bf y}}
\newcommand{\hby}{\hat{\bf y}}
\newcommand{\br}{{\bf r}}
\newcommand{\hbr}{\hat{\bf r}}
\newcommand{\bj}{{\bf j}}
\newcommand{\bk}{{\bf k}}
\newcommand{\bn}{{\bf n}}
\newcommand{\bv}{{\bf v}}
\newcommand{\bp}{{\bf p}}
\newcommand{\bq}{{\bf q}}
\newcommand{\tp}{\tilde{p}}
\newcommand{\tbp}{\tilde{\bf p}}
\newcommand{\bu}{{\bf u}}
\newcommand{\hbz}{\hat{\bf z}}
\newcommand{\bA}{{\bf A}}
\newcommand{\calA}{\mathcal{A}}
\newcommand{\calB}{\mathcal{B}}
\newcommand{\tC}{\tilde{C}}
\newcommand{\bD}{{\bf D}}
\newcommand{\bE}{{\bf E}}
\newcommand{\calF}{\mathcal{F}}
\newcommand{\bB}{{\bf B}}
\newcommand{\bG}{{\bf G}}
\newcommand{\calG}{\mathcal{G}}
\newcommand{\obG}{\overleftrightarrow{\bf G}}
\newcommand{\bJ}{{\bf J}}
\newcommand{\bK}{{\bf K}}
\newcommand{\bL}{{\bf L}}
\newcommand{\tL}{\tilde{L}}
\newcommand{\bP}{{\bf P}}
\newcommand{\calP}{\mathcal{P}}
\newcommand{\bQ}{{\bf Q}}
\newcommand{\bR}{{\bf R}}
\newcommand{\bS}{{\bf S}}
\newcommand{\bH}{{\bf H}}
\newcommand{\balpha}{\mbox{\boldmath $\alpha$}}
\newcommand{\talpha}{\tilde{\alpha}}
\newcommand{\bsigma}{\mbox{\boldmath $\sigma$}}
\newcommand{\hbeta}{\hat{\mbox{\boldmath $\eta$}}}
\newcommand{\bSigma}{\mbox{\boldmath $\Sigma$}}
\newcommand{\bomega}{\mbox{\boldmath $\omega$}}
\newcommand{\bpi}{\mbox{\boldmath $\pi$}}
\newcommand{\bphi}{\mbox{\boldmath $\phi$}}
\newcommand{\hbphi}{\hat{\mbox{\boldmath $\phi$}}}
\newcommand{\btheta}{\mbox{\boldmath $\theta$}}
\newcommand{\hbtheta}{\hat{\mbox{\boldmath $\theta$}}}
\newcommand{\hbxi}{\hat{\mbox{\boldmath $\xi$}}}
\newcommand{\hbzeta}{\hat{\mbox{\boldmath $\zeta$}}}
\newcommand{\brho}{\mbox{\boldmath $\rho$}}
\newcommand{\bnabla}{\mbox{\boldmath $\nabla$}}
\newcommand{\bmu}{\mbox{\boldmath $\mu$}}
\newcommand{\bepsilon}{\mbox{\boldmath $\epsilon$}}

\newcommand{\iLambda}{{\it \Lambda}}
\newcommand{\cL}{{\cal L}}
\newcommand{\cH}{{\cal H}}
\newcommand{\cU}{{\cal U}}
\newcommand{\cT}{{\cal T}}

\newcommand{\be}{\begin{equation}}
\newcommand{\ee}{\end{equation}}
\newcommand{\bea}{\begin{eqnarray}}
\newcommand{\eea}{\end{eqnarray}}
\newcommand{\beqa}{\begin{eqnarray*}}
\newcommand{\eeqa}{\end{eqnarray*}}
\newcommand{\nn}{\nonumber}
\newcommand{\DD}{\displaystyle}

\newcommand{\ba}{\begin{array}{c}}
\newcommand{\baa}{\begin{array}{cc}}
\newcommand{\baaa}{\begin{array}{ccc}}
\newcommand{\baaaa}{\begin{array}{cccc}}
\newcommand{\ea}{\end{array}}

\newcommand{\bma}{\left[\begin{array}{c}}
\newcommand{\bmaa}{\left[\begin{array}{cc}}
\newcommand{\bmaaa}{\left[\begin{array}{ccc}}
\newcommand{\bmaaaa}{\left[\begin{array}{cccc}}
\newcommand{\ema}{\end{array}\right]}

\preprint{AIP/123-QED}

\title{Linear and nonlinear optical response based on many-body $GW$-Bethe-Salpeter and Kadanoff-Baym approaches for two-dimensional layered semiconductors}

\author{Dmitry Skachkov$^{(1)}$}
 \email{dmitry.skachkoff@ucf.edu}j

\author{Dirk R. Englund$^{(2)}$}
\email{englund@mit.edu}

\author{Michael N. Leuenberger$^{(1,3)}$}

\email{michael.leuenberger@ucf.edu}

\affiliation{$^{(1)}$ NanoScience Technology Center and Department of Physics, University of Central Florida, Orlando, FL 32826, USA. \\
$^{(2)}$ Department of Electrical Engineering and Computer Science, Massachusetts Institute of Technology, Cambridge, MA 02139, USA.\\
$^{(3)}$ College of Optics and Photonics, University of Central Florida, Orlando, FL 32826, USA.}

\date{\today}

\begin{abstract}
The family of 2D layered semiconductors, including transition metal chalcogenides (TMCs) of the form MX (M=Ga, In; X=S, Se, Te) exhibit exceptional nonlinear optical properties. 
The energetically most favorable crystal ordering for nonlinear response is the AB layer stacking, which breaks central inversion symmetry for an arbitrary number of layers, resulting in non-zero off-diagonal elements of the $\chi^{(2n')}$ tensor, $n'$ being a positive integer, for arbitrary thickness of the materials.  
We perform first-principles many-body calculations of bandstructures and linear and nonlinear optical responses of monolayer (ML) and bulk TMC crystals based on $GW$-Bethe-Salpeter and Kadanoff-Baym approaches in and out of equilibrium, respectively, while taking many-body band gap renormalization and excitonic effects into account. We develop a detailed analysis of the linear and nonlinear optical selection rules by means of group and representation theory, showing strong connection to crystal symmetry and orbital characters of the bands and providing a method to predict the strength of linear and nonlinear response of new materials. We observe the general trend that the lowest-energy excitons are dark in 2D ML TMCs whereas they are bright or mixed bright-dark in 3D bulk TMCs, which we attribute to the difference between spatially dependent screening in 2D and constant screening in 3D. In particular, we derive general formulas for the nonlinear optical response based on exciton states in semiconductor materials. We find anti-bound excitons in ML GaS, which we attribute to dominant exciton exchange interaction. We show that by choosing elements with larger mass and by reducing the detuning energy it is possible to increase the nonlinear response not only for $\chi^{(2)}$ and $\chi^{(3)}$, responsible for SHG and third harmonic generation (THG), but also in general for $\chi^{(n)}$ nonlinear response with $n>3$, giving rise to high harmonic generation (HHG) in 2D semiconductor materials. We achieve good to excellent agreement with experimental measurements of linear exciton spectra and nonlinear coefficients for ML and bulk GaS, GaSe, and GaTe for $\chi^{(n)}$, $n=2$ and $n=3$. We predict the nonlinear response of InTe and AlTe for $2<n<7$. The HHG regime has so far been rarely considered, both theoretically and experimentally. We predict high values of SHG, THG, and HHG for the TMC family of materials for $2\le n\le 7$, which is in agreement with similar trends in transition metal dichalcogenides. 
\end{abstract}

\maketitle

\section{Introduction}

Layered 2D semiconductors have attracted a lot of attention due to their intriguing electronic and optical properties with a wide range of promising applications \cite{Thakur2024,Lemme2022,Burmeister2021,Lu2020,Arora2021,Bernardi2017,Zhu2023,Kumbhakar2021,Huang2022,Shanmugam2022,Kumbhakar2023,Lee2016,Batool2023,Xu2021,Yongzhuo2020,Li_REVIEW_2018}, such as
in high-performance low-power electronic components \cite{Huang2022,Nandan2023,Kanungo2022,Meena2023,Sangwan2018}, catalysis \cite{Cho2024,Wang2022}, ultrafast non-volatile flash memory \cite{Liu2021,Wu2021}, logic-in-memory \cite{Marega2020}, and reconfigurable logic elements \cite{Pan2020,PWu2021}.  
2D semiconductors are also a platform for self-assembly of nanostructures \cite{Hu2023}, to create ideal Schottky contact \cite{Wang2024_2,Zhang2021,Skachkov2021}, and for generation of intense electric fields \cite{Weintrub2022}. 
The applications of 2D semiconductors in optics include photodetectors \cite{Guo2024,Wang2024}, optical modulators \cite{Sun2016}, nonlinear optical rectification \cite{Taghizadeh2021,Shi2023}, 2D semiconductor lasers \cite{Xiao2022}, light emitting diodes (LEDs) \cite{Ahmed2023} including quantum dot LEDs \cite{Jung2023}, second harmonic generation (SHG) \cite{Hung2024,Huang2024}, third harmonic generation (THG) \cite{Nejad2024,Popkova2021}, high-harmonic generation (HHG) \cite{Lee2024,Yue2022}, direct current generation from SHG \cite{Zhang2019,Akamatsu2021,Sotome2021}, shift current generation due to large excitonic effects \cite{Akamatsu2021,Chan2021}, photovoltaics \cite{Ren2024,Li2023,Aryal2022}, plasmonics \cite{Agarwal2018}, excitonics \cite{Du2024}, valleytronics \cite{Yu2015}, and twistable electronics, called twistronics \cite{Ciarrocchi2022,Li2024,Carr2017,Hennighausen2021,Gupta2024,RIBEIROPALAU2018}.

Monolayer (ML) transition-metal chalcogenides (TMCs) are direct band gap semiconductors \cite{Direct_Band_Gap_1,Direct_Band_gap_2,review_TMDCS,review_TMDCS_2}, which can be used to produce smaller and more energy efficient devices, such as transistors and integrated circuits. In the ideal case, the 2D structure is free of any dangling bonds, resulting in clean 2D surfaces, even for an arbitrary number of 2D layers. 
Moreover, their band gaps lie in the visible region, which makes them highly responsive when exposed to visible light, a property with potential applications in optical photodetection. 
The binary IIA-VIA group compounds, like Gallium Selenide (GaSe), have attracted a lot of attention due to distinctive optoelectronic properties. While GaSe is one of the most promising nonlinear-optical materials for infrared (IR) frequency conversion, SHG, and THG, the theoretical aspects in terms of many-body excitonic response of TMCs have not been studied so far. 

The strength of the nonlinear optical response depends on the symmetry of the 2D layered semiconductor materials. 
In a ML the inversion symmetry can be broken due to specific geometry of the system, which increases the nonlinear optical response. 
Examples of such systems are $\epsilon$-GaSe \cite{Zhou2015}, $\epsilon$-GeSe \cite{Wang2017}, CdS \cite{Ren2018}, and $\beta$-NP \cite{Kolos2021} crystals, which all exhibit AB stacking of their 2D MLs. Experimentally, the second harmonic generation (SHG) in GaSe is the strongest among all the 2D layered crystals.
In general, different stacking orders between layers affect the band gap type and the band energies \cite{Lai2023}; in order to preserve the even-order $\chi^{(2n')}$, $n'$ being a positive integer, nonlinear responses, it is necessary to avoid restoring the inversion symmetry during stacking. 
This is possible to achieve, for example, by means of AB stacking.   
In order to avoid losses and decoherence, the nonlinear interactions between the photons must occur via virtual quantum states inside the nonlinear materials, which means that the photon energy is tuned below the band gap of the nonlinear material.  

Here, we focus on the nonlinear interaction due to excitons in TMCs. We use first-principles methods to calculate the linear (LR) and the nonlinear responses (NLR) of the electronic system under optical excitation. To make our paper transparent to the reader, we provide reviews of the used theoretical methods, which are the Bethe-Salpeter equation (BSE), Hedin's equations, the $GW$ approximation, the COHSEX approximation, and the Kadanoff-Baym equations (KBEs).


We consider the TMCs of the form MX (M=Ga, In; X=S, Se, Te).
Here we show that using the simple rule of substituting lighter elements by more heavy elements from the same column of the Periodic Table, we show that it is possible to increase the off-resonance NLR for $\chi^{(n)}$, $n$ being a positive integer.
The observed effect can be understood using an empirical model based on the many-body perturbation theory with the electric dipole formalism in the length gauge.

The rest of the paper is organized as follows. 
Sec.~\ref{sec:theory} analyzes and reviews first-principle approaches for describing LRs and NLRs of semiconductor materials including excitonic effects. We describe in detail the $GW$ approximation for obtaining the self-energies of the Bloch states for achieving good agreement with experimental data, especially the band gap renormalization of semiconductor materials, the COHSEX approximation of the self-energy, the derivation of the BSE for equilibrium calculations starting from Hedin's equations and its connection to the Wannier equation for anisotropic excitons, the LR due to excitons, the real-time Kadanoff-Baym (KB) approach for out-of-equilibrium dynamics using Keldysh Green's functions for analyzing LR and NLR to arbitrary order $\chi^{(n)}$, $n$ being a positive integer, and the dynamical polarization including NLR and its connection to topological quantities, such as Berry connections and Berry curvatures.  In the subsections we provide detailed analysis of the linear and nonlinear optical selection rules by means of group and representation theory. In particular, we derive general formulas for the nonlinear optical response based on exciton states in semiconductor materials. These formulas provide a simple interpretation of the linear and nonlinear optical selection rules in terms of electric dipole matrix elements between Bloch states and electric dipole matrix elements between the relative-motion eigenstates of excitons.
Sec.~\ref{sec:numerics} describes all steps of the numerical procedures and the technical details of the ab-initio calculations based on the QUANTUM ESPRESSO and YAMBO software packages, where we show that good to excellent agreement with experimental data can be achieved by using the ev$GW$ method.
Sec.~\ref{sec:optical} presents the results for LR and NLR parameters for TMCs.       
We find that the LRs obtained from the BSE and the Kadanoff-Baym equation (KBE) are nearly identical and agree well with experimental data. We also find that the NLR obtained for some TMCs using the KBE also agrees well with available experimental data.

 \section{Theoretical Methods}
 \label{sec:theory}
 In this section we consider first-principles approaches combined with many-particle quantum field theory to describe excitons in 2D and 3D semiconductor materials. 
 For a clear understanding of the theoretical methods applied to our numerical calculations and for convenience, we review the $GW$ approximation to the self-energy, the COHSEX approximation to the $GW$ self-energy, the BSE solver for LR of equilibrium states, and NLR by means of the KBEs describing non-equilibrium dynamics, all based on the Kohn-Sham states obtained by means of density functional theory (DFT).

\subsection{$GW$ approach}

\begin{figure}
    \centering
    \includegraphics[width=3.5in]{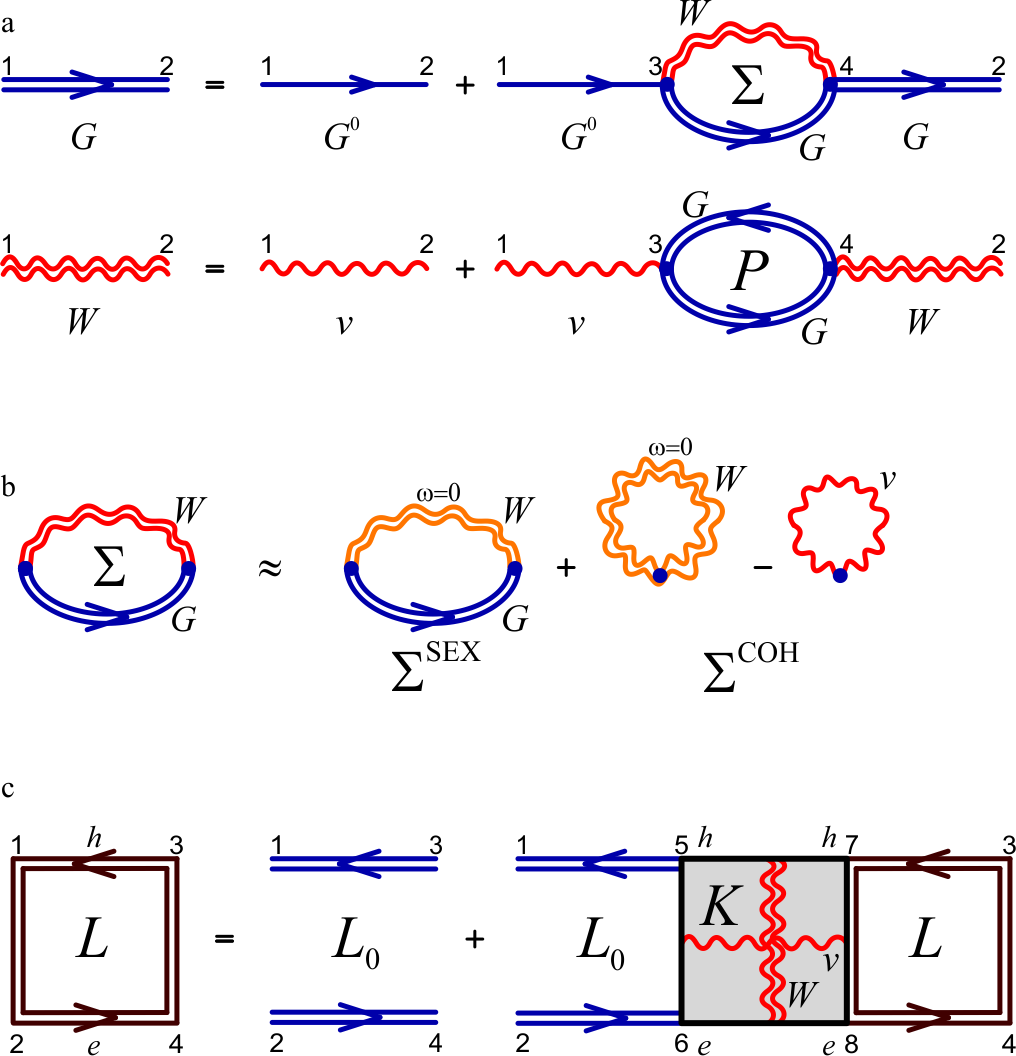}
    \caption{Feynman diagrams for (a) $GW$ method, Eq. \eqref{eq:GW1}-\eqref{eq:GW2}, (b) COHSEX approximation to self-energy, Eq.~\eqref{eq:SEX}-\eqref{eq:COH}, and (c) BSE, Eq.~\eqref{BSE_LGW}.}
    \vspace{5mm}
\label{fig:Feynman_diagrams}
\end{figure}

In order to correct the underestimated band gap obtained in DFT, we use the many-body $GW$ approach that accounts for electron self-interaction.
The main idea is to approximate the self-energy in the first of Hedin's equations 
\cite{Hedin1999,Onida2002}.

The first Hedin equation is the Dyson equation connecting the interacting Green's function $G$, the non-interacting Green's function $G^0$, and the self-energy $\Sigma$, i.e.
\begin{equation}
\label{eq:Hedin2}
G(12) = G^0(12)+G^0(13)\Sigma(34)G(42).  
\end{equation}
The numbers in Eq.~\eqref{eq:Hedin2} are combinations of the particles' parameters, for example, $1 = \left(\mathbf{r}_1, t_1 \right)$ in real space, or $1 = \left(\mathbf{k}_1, \omega_1 \right)$ in reciprocal space, where $\mathbf{r}_1$ is the position of electron 1 with spin $\xi_1$ at time $t_1$.   
The remaining Hedin equations are
\begin{eqnarray}
\Sigma(12) &=& i G(13) \Gamma(324) W(41),  \label{eq:Hedin1}\\
\Gamma(123) &=& \delta(12) \delta(13), \label{eq:Hedin3}\\
& &+ \frac{\delta \Sigma(12)}{\delta G(45)} G(46) G(75) \Gamma(673),  \nonumber\\
P(12) &=& -i G(13) G(41) \Gamma(342),  \label{eq:Hedin4}\\
W(12) &=& v(12)+ v(13) P(34) W(42). \label{eq:Hedin5}
\end{eqnarray}
In random phase approximation (RPA) \cite{Bohm1953} we neglect vertex corrections, i.e. by setting $\Gamma(123)=\delta(12) \delta(13)$. In this case, the irreducible polarization  is determined by the non-interacting quasielectron-quasihole pairs, i.e.
\begin{equation}
\label{eq:GW_P}
P_{GW}(12)=-i G(12) G\left(21^{+}\right),
\end{equation}
and the self-energy is
\begin{equation}
\label{eq:GW_S}
\Sigma_{GW}(12)=i G(12) W(21),
\end{equation}
which contains the interacting Green's function $G$ and the dynamically screened Coulomb potential $W$, what gives the name of the method.  

Thus, the $GW$ method is the first iteration in solving Hedin's equations iteratively, and it consists of two connected Dyson equations for one-particle (two-point) functions $G$ and $W$, i.e.
\begin{eqnarray}
G(12) &=& G^0(12) + G^0(13) \Sigma_{GW}(34) G(42)   \label{eq:GW1} \\
W(12) &=& v(12) + v(13) P_{GW}(34) W(42) \label{eq:GW2} 
\end{eqnarray}
The Feynman diagram for the $GW$ self-energy, representing the many-body energy renormalization to quasi-particles, is shown in Fig.~\ref{fig:Feynman_diagrams}. The non-interacting Green's function $G^0$ is represented by thin blue lines, whereas the interacting Green's function $G$ is shown by double (connected) blue lines. The same notation is used for $W$ and $v$, where the red double wiggly lines represent the dynamically screened Coulomb interaction $W$, and the red single wiggly lines represent the bare Coulomb interaction $v$. The upper part of the figure on Fig.~\ref{fig:Feynman_diagrams}a shows the Dyson equation \eqref{eq:GW1} including Eq.~\eqref{eq:GW_S} for $\Sigma$. The lower part shows the Dyson equation \eqref{eq:GW2} including Eq.~\eqref{eq:GW_P} for the polarization $P$.

\subsection{COHSEX approximation}


Calculating the $GW$ self-energy is computationally expensive. In order to develop a more efficient way to calculate the self-energy for some particular cases, Hedin proposed the static retarded approach, i.e. the Coulomb-hole screened-exchange (COHSEX) approximation to the self-energy \cite{Berger2021,Kang2010}. 
We can rewrite the self-energy in Eq.~\eqref{eq:GW_S} as
\begin{eqnarray}
\label{eq:S2}
\Sigma = iGW &=& iGv + iGW_p = \nonumber  \\
             &=& \Sigma_x + \Sigma_c,
\end{eqnarray}
where $W_p = W - v$. The left part in Eq.~\eqref{eq:S2} is the exchange term, and the right part is the correlation term of the self-energy. For the static case ($\omega=0$) we can expand the Green function in the correlation part at $\tau=0$ around the one-half of $(\tau+\eta)$ and $(\tau-\eta)$ points. Then the correlation part splits into two parts. One of these parts in combination with the exchange term $\Sigma_x$ gives the screened exchange (SEX) term (for derivation see Ref.~\cite{Berger2021}). The remaining part of $\Sigma_c$ forms the Coulomb-hole (COH) term, which represents the interaction of the electron with changes in the potential due to screening, i.e. 
\begin{eqnarray}
\Sigma^{\text{SEX}} (12) &=& iG(12) W(21)\bigr|_{\omega=0}  \label{eq:SEX} \\
\Sigma^{\text{COH}} (12) &=& \frac{1}{2} \delta(12) W_p(21)\bigr|_{\omega=0}    \label{eq:COH} 
\end{eqnarray}
This approach can be considered as an extension of the Hartree method with static screening. The Feynman diagram for the COHSEX method is shown in Fig.~\ref{fig:Feynman_diagrams}b. The statically screened Coulomb interaction $W$ is denoted by orange double wiggly lines. 

\subsection{Bethe-Salpeter Approach}



In order to account for exciton bound states of electron-hole pairs due to the Coulomb interaction in equilibrium, we use the Bethe-Salpeter approach \cite{Strinati1988,Bussi2004}.  

The BSE can be obtained by iterating twice over Hedin's equations (\ref{eq:Hedin2}) - (\ref{eq:Hedin5}) \cite{Onida2002}. This means in the first iteration one sets $\Sigma=0$ and $\Gamma=\delta\delta$, resulting in $\Sigma=0\rightarrow G=G_0\rightarrow\Gamma=\delta\delta\rightarrow P=-iG_0G_0\rightarrow W=v+vPW$.
In the second iteration the approximation $\frac{\delta \Sigma(12)}{\delta G(45)}=iW$ is used, giving
$\Sigma=iGW\rightarrow \Gamma=\delta\delta+iWGG\Gamma$, which is the vertex equation. 
By multiplying this vertex equation from the left with $-iGG$, integrating, and defining the generalized three-point polarizability $^3P=-iGG\Gamma$, one can write the Dyson-like equation
\begin{equation}
^3P=-iGG+iGGW\;^3P.
\end{equation}
Introducing the reducible polarizability $L_0(1234)=-iG(13)G(42)$ in terms of the full Green's function $G$
and extending the 3-point irreducible polarizability $^3P(123)$ to a 4-point irreducible polarizability $^4P(1234)$, one obtains
\begin{equation}
\label{BSE_4P}
^4P(1234)=L_0(1234)-L_0(1256)^4W(5678) {^4P(7834)},
\end{equation}
where $^4W(1,2,3,4)=W(1,2)\delta(1,3)\delta(2,4)$.
Then one can take advantage of the relation between the 4-point reducible polarizability $L(1234)$ and $^4P(1234)$, which is
\begin{equation}
\label{BSE_L}
L(1234)=^4P(1234)+^4P(1256)^4v(5678) L(7834),
\end{equation}
where $^4v(1,2,3,4)=v(1,3)\delta(1,2)\delta(3,4)$.
Inserting Eq.~(\ref{BSE_4P}) into Eq.~(\ref{BSE_L}), one gets the BSE in $GW$ approximation, i.e.
\begin{equation}
\label{BSE_LGW}
L(1234)=L_0(1234)+L_0(1256)K(5678) L(7834).
\end{equation}
If the full Green's function $G$ is approximated by $G_0$, then the BSE in $G_0W$ approximation is
\begin{equation}
\label{BSE_ChiG0W}
\chi(1234)=\chi_0(1234)+\chi_0(1256)K(5678) \chi(7834).
\end{equation}
The BSE describes many-body two-particle interactions \cite{Onida2002}, for example between electrons and holes that form bound exciton states. The definitions of the polarizabilities and susceptibilites are given in Sec.~I of the Supplementary Information.
The kernel $K$ contains the repulsive electron-hole exchange interaction $\bar{v}$ and the attractive electron-hole interaction $W$, i.e. 
\begin{equation}
K\left(1234\right)=\delta \left(12\right)\delta \left(34\right)\bar{v}\left(13\right)-\delta \left(13\right)\delta \left(24\right)W\left(12\right)
\end{equation}
The BSE and its kernel $K$ are shown in Fig.~\ref{fig:Feynman_diagrams}c. The kernel $K$ contains the attractive direct interaction 5-6 ($e-h$) and 7-8 ($e-h$) (vertical double wiggly red lines) and the repulsive exchange interaction 5-7 ($h-h$) and 6-8 ($e-e$) (horizontal single wiggly red line).

Since any four-point function, in particular the polarizability $L$, can be written as \cite{Onida2002}
\begin{equation}
\begin{aligned}
& L\left(\mathbf{r}_1, \mathbf{r}_1^{\prime} ; \mathbf{r}_2, \mathbf{r}_2^{\prime}\right) \\
& \quad=\sum_{n_1 \cdots n_4} \psi_{n_1}^*\left(\mathbf{r}_1\right) \psi_{n_2}\left(\mathbf{r}_1^{\prime}\right) \psi_{n_3}\left(\mathbf{r}_2\right) \psi_{n_4}^*\left(\mathbf{r}_2^{\prime}\right)\; L_{\left(n_1 n_2\right)\left(n_3 n_4\right)}
\end{aligned}
\end{equation}
in terms of the single-particle Bloch functions $\psi_{n,\bk}\left(\mathbf{r}\right)$, where $n$ is the band index and $\bk$ the wave vector absorbed in the band index $n_i,\bk_i\rightarrow n_i$, it is possible to represent the BSE (\ref{BSE_LGW}) in the band index and momentum space as
\be
\begin{aligned}
L_{(n_1 n_2)(n_3 n_4)}(\omega)= & L^0_{n_1 n_2}(\omega)\left[\delta_{n_1 n_3} \delta_{n_2 n_4}\right. \\
& +\left.\sum_{n_5 n_6} K_{\substack{n_1 n_2 \\ n_5 n_6}}(\omega) L_{(n_5 n_6)(n_3 n_4)}(\omega)\right],
\end{aligned}
\ee
where
\be
L^0_{(n_1 n_2)(n_3 n_4)}(\omega)=\frac{f_{n_2}-f_{n_1}}{\epsilon_{n_2}-\epsilon_{n_1}-\omega} \delta_{n_1, n_3} \delta_{n_2, n_4}
\label{eq:P0}
\ee
is diagonal.
The BSE can be rewritten as
\be
\begin{aligned}
& \sum_{n_5 n_6}\left[\delta_{n_1 n_5} \delta_{n_2 n_6}-L_{n_1 n_2}^0(\omega) K_{\substack{n_1 n_2 \\ n_5 n_6}}(\omega)\right]\;L_{(n_5 n_6)(n_3 n_4)}(\omega) \\
& =L_{n_1 n_2}^0(\omega),
\end{aligned}
\ee
and then, using Eq.~(\ref{eq:P0}), recast into an eigenvalue equation
\be
\begin{aligned}
& \sum_{n_5 n_6}\left[\left(\epsilon_{n_2}-\epsilon_{n_1}-\omega\right) \delta_{n_1 n_5} \delta_{n_2 n_6}-\left(f_{n_2}-f_{n_1}\right) K_{\substack{n_1 n_2 \\ n_5 n_6}}(\omega)\right] \\
& \times\;L_{\substack{n_5 n_6 \\ n_3 n_4}}(\omega)=f_{n_2}-f_{n_1},
\end{aligned}
\label{eq:BSE_eigenvalue}
\ee
for which the effective two-particle Hamiltonian
\begin{equation}
\begin{aligned}
H_{\substack{n_1\bk_1 n_2,\bk_2 \\ n_5\bk_5 n_6,\bk_6}}^{2 p} \equiv & \left(\epsilon_{n_2,\bk_2}-\epsilon_{n_1\bk_1}\right) \delta_{n_1, n_5} \delta_{n_2, n_6}\delta_{\bk_1\bk_5}\delta_{\bk_2\bk_6} \\
& -\left(f_{n_1,\bk_1}-f_{n_2,\bk_2}\right) K_{\substack{n_1\bk_1 n_2,\bk_2 \\ n_5\bk_5 n_6,\bk_6}}
\end{aligned}
\label{eq:H2p}
\end{equation}
can be defined.
The solution of the BSE in the band index space is
\be
L_{\left(n_1 n_2\right)\left(n_3 n_4\right)}=\left[H^{2 p}-I \omega\right]_{\left(n_1 n_2\right)\left(n_3 n_4\right)}^{-1}\left(f_{n_4}-f_{n_3}\right).
\label{eq:BSE_solution}
\ee
This solution has poles at the eigenenergies of the effective two-particle Hamiltonian in Eq.~(\ref{eq:H2p}). 

In the Tamm-Dancoff approximation the off-diagonal coupling block is neglected, and therefore only the upper diagonal resonant block in the hole-particle subspace $\{vc\}$ is needed \cite{Onida2002}, i.e.
\begin{equation}
H_{(v c)\left(v^{\prime} c^{\prime}\right)}^{2p,res} \equiv\left(\epsilon_c-\epsilon_v\right) \delta_{v, v^{\prime}} \delta_{c, c^{\prime}}+\left(f_v-f_c\right)K_{(v c)\left(v^{\prime} c^{\prime}\right)} .
\label{H_BSE}
\end{equation}
Thus, one obtains the effective time-independent Schr\"odinger equation
\begin{equation}
H^{2p,res} A_{\lambda,\bk_e,\bk_h}^{vc}=E_{\lambda,\bk_e,\bk_h} A_{\lambda,\bk_e,\bk_h}^{vc} 
\label{H_2p}
\end{equation}
for excitons with eigenvectors $A_{\lambda,\bk_e,\bk_h}$ and their energy eigenvalues $E_{\lambda,\bk_e,\bk_h}$, where $\lambda$ denotes the relative-motion (hydrogen-like $\lambda=1s, 2s, 2p, \ldots$ in 3D) quantum number of the relative motion of the electron and hole. The exciton wavefunction can then be written as \cite{Yu&Cardona}
\be
\Psi_{\bK\lambda}\left(\br_e,\br_h\right)=\sum_{\bk_e,\bk_h} A_{\lambda,\bk_e,\bk_h}^{vc} \psi_{\bk_e}\left(\br_e\right) \psi_{\bk_h}\left(\br_h\right),
\ee
where $\bK$ is the center-of-mass momentum, and $\psi_{\bk_e}\left(\br_e\right)$ and $\psi_{\bk_h}\left(\br_h\right)$ are the Bloch functions of the electron and hole, respectively. Since the electron and hole are localized relative to their center-of-mass motion, it is useful to express the exciton wavefunction in terms of electron and hole Wannier functions $w_{\bR_e}\left(\br_e\right)$ and $w_{\bR_h}\left(\br_h\right)$, respectively, by means of the Fourier transformation
\be
\left. \left\langle \br \right|n\bR \right\rangle  \equiv w_{n\bR}(\br) = \frac{1}{{\sqrt N }}\sum\limits_{\bf{k}} {{e^{ - i{\bf{k}}\cdot{\bf{R}}}}} {\psi _{n{\bf{k}}}}({\bf{r}}).
\ee
Then, one obtains
\be
\Psi_{\bK\lambda}\left(\br_e,\br_h\right)=\sum_{\bR_e,\bR_h} \Phi_{\bK\lambda,\bR_e,\bR_h}^{vc} w_{\bR_e}\left(\br_e\right) w_{\bR_h}\left(\br_h\right),
\ee
where 
\be
\Phi_{\bK\lambda,\bR_e,\bR_h}^{vc}=\frac{1}{N}\sum\limits_{{{\bf{k}}_e},{{\bf{k}}_h}} {A_{\bK\lambda ,{{\bf{k}}_e},{{\bf{k}}_h}}^{vc}{e^{i{{\bf{k}}_e}\cdot{{\bf{R}}_e}}}{e^{i{{\bf{k}}_h}\cdot{{\bf{R}}_h}}}}
\ee
is the exciton envelope wavefunction.
The time-independent Schr\"odinger equation for this exciton wavefunction is
\begin{align}
& {\left[-\left(\frac{\hbar^2}{2 m_{\mathrm{e}}}\right) \nabla_{\boldsymbol{R}_{\mathrm{e}}}^2-\left(\frac{\hbar^2}{2 m_{\mathrm{h}}}\right) \nabla_{\boldsymbol{R}_{\mathrm{h}}}^2+V(r)\right] \Phi_{\bK\lambda}\left(\bR_e, \bR_h\right)} \nn\\
& \quad =E_\lambda \Phi_{\bK\lambda}\left(\bR_e, \bR_h\right),
\end{align}
In terms of the center-of-mass coordinate 
$\bR=\frac{m_e \bR_e+m_h \bR_h}{m_e+m_h}$
and relative coordinate $\br=\bR_e-\bR_h$, along with a product ansatz 
$\Phi_{\bK\lambda}^{vc}\left(\bR_e, \bR_h\right)=\xi_\bK(\bR)\phi_\lambda(\br)$, one obtains two decoupled time-independent Schr\"odinger equations
\begin{align}
& -\frac{\hbar^2}{2 M} \nabla_{\bR}^2 \xi_\bK(\bR)=E_\bK \xi_\bK(\bR), \\
& \left(-\frac{\hbar^2}{2 \mu} \nabla_{\boldsymbol{r}}^2+V(r)\right)\phi_\lambda(\br)=E_\lambda \phi_\lambda(\br),
\end{align}
for the center-of-mass motion and the relative motion, respectively. $\mu$ is the reduced mass of the exciton with $\frac{1}{\mu}=\frac{1}{m_{\mathrm{e}}}+\frac{1}{m_{\mathrm{h}}}$ and $M=m_e+m_h$.
The center-of-mass eigenfunction is
\be
\xi_\bK(\bR)=\frac{1}{\sqrt{N}}e^{i\bK\cdot\bR}.
\ee
The exciton wavefunction can now be written in the form
\be
\Psi_{\bK\lambda}\left(\bR,r\right)=(1/\sqrt{N}) {\phi _\lambda }(r){e^{ - i{\bf{k}}\cdot{\bf{r}}}}{\psi _{{{\bf{k}}_e}}}({{\bf{r}}_e}){\psi _{{{\bf{k}}_h}}}({{\bf{r}}_h}),
\ee
In the electric dipole approximation, we can neglect the momentum of the photon, thereby setting $\bK\approx 0$.
Then we can write
\be
\Psi_{\bK=0\lambda}\left(\bR,r\right)=(1/\sqrt{N})\phi_\lambda(r)e^{-i{\bf{k}}\cdot{\bf{r}}}\psi_\bk(\br_e)\psi_{-\bk}(\br_h).
\label{eq:exciton}
\ee
This form allows us to identify the optical selection rules for LR and NLR of excitons in terms of Bloch states and relative-motion states in Sec.~\ref{sec:selection_rules_excitons}.
The exciton binding energy for the 3D isotropic case is \cite{Yu&Cardona}
\be
E_{nm}^{\rm 3D, iso}=-\frac{\mu e^4}{2\hbar^2\varepsilon^2n^2}.
\ee
The kinetic energy of the center-of-mass motion is $E_{\rm COM}=\hbar^2K^2/2M$.

\subsection{Excitons in TMCs}
\label{sec:excitons_GaS}
The TMCs in bulk form host anisotropic excitons. Their relative-motion Hamiltonian reads
\be
\tilde{H}=-\frac{\hbar^2}{2 \mu_{\perp}}\left(\partial_x^2+\partial_y^2\right)-\frac{\hbar^2}{2 \mu_{\|}} \partial_z^2-\frac{e^2}{\sqrt{\varepsilon_{\perp} \varepsilon_{\|}} \sqrt{x^2+y^2+\left(\frac{\varepsilon_{\perp}}{\varepsilon_{\mid |}}\right) z^2}}.
\label{eq:exciton_GaSe}
\ee
The indices $\perp$ and $\|$ identify the directions perpendicular to the crystal $c$-axis, corresponding to the $xy$-plane, and parallel to the $c$-axis, corresponding to the $z$-direction. The $\mu_{\perp}, \mu_{\|}$are the in-plane and out-of-plane reduced effective masses, and $\varepsilon_{\perp}, \varepsilon_{\|}$ are the respective dielectric constants.

Analytical eigenstates and eigenenergies of $\tilde{H}$ have not been found. However, several approximate theoretical models have been derived \cite{Deverin1969,Baldereschi1970,Gerlach1975}. The first step is to transform to new coordinates $x'=x$, $y'=y$, and $z'=\sqrt{\mu_{\|}^* / \mu_{\perp}^*} z$. Then the transformed Hamiltonian reads
\be
\tilde{H}^{\prime}=-\frac{h^2}{2 \mu_{\perp}}\left(\partial_{x'}^2+\partial_{y'}^2+\partial_{z'}^2\right)-\frac{e^2}{\sqrt{\varepsilon_{\perp} \varepsilon_{\|}} \sqrt{{x'}^2+{y'}^2+\gamma {z'}^2}},
\ee
where the anisotropy parameter $\gamma=\mu_{\perp} \varepsilon_{\perp} / \mu_{\|} \varepsilon_{||}$ has been defined \cite{Deverin1969,Baldereschi1970}. 
In terms of modified Rydberg units $E_0=e^4 \mu_{\perp} / 2 \hbar^2 \varepsilon_{\perp} \varepsilon_{\|}$ and Bohr radii $a_0=\hbar^2 \sqrt{\varepsilon_{\perp} \varepsilon_1} / \mu_{\perp} e^2$ the Schr\"odinger equation reads
\be
\left(-\bnabla_{\br'}^2-\frac{2}{\sqrt{{x'}^2+{y'}^2+\gamma {z'}^2}}\right) \psi(\br')=E \psi(\br')
\ee
Alternatively, it is possible to define the anisotropy parameter $\alpha=(1-\gamma)$ \cite{Gerlach1975}. Then the Coulomb potential is expressed as
\be
\frac{2}{\sqrt{x^2+y^2}+\overline{A z^2}}=\frac{2}{r} \frac{1}{\sqrt{1-\alpha \cos ^2 \vartheta}}=\frac{2}{r} f(\alpha, \vartheta),
\ee
where the function $f(\alpha, \vartheta)=1/\sqrt{1-\alpha \cos ^2 \vartheta}$ describes the deviation of the Coulomb interaction from the isotropic case. 
For $\alpha=0$ ($\gamma=1$) the Coulomb potential is isotropic. For $\alpha=1$ ($\gamma=0$) the two-dimensional hydrogen potential is obtained. 
In the case of GaSe, the dielectric constants have been experimentally determined to be $\varepsilon_{\perp}=10.4$ and $\varepsilon_{\|}=6.0$ in Ref.~\cite{Toullec1980}, and $\varepsilon_{\perp}=10.2$ and $\varepsilon_{\|}=7.9$ in Ref.~\cite{Leung1966}.
The ratio of the in-plane and out-of-plane effective masses has been experimentally determined to be $\mu_{\perp}/\mu_{\|}=1.19\pm 0.07$ in Ref.~\cite{Toullec1980}, $\mu_{\perp}/\mu_{\|}=1\pm 0.2$ in Ref.~\cite{Mooser1973}, and $\mu_{\perp}/\mu_{\|}=1.17$ in Ref.~\cite{Ottaviani1974}.
Thus, using the values in Ref.~\cite{Toullec1980} the anisotropy parameter for GaSe is $\gamma\approx 2$.
Using an averaging technique along with perturbation theory, Ref.~\cite{Gerlach1975} obtains the eigenfunctions
\be
\phi_{(n l) m}(\boldsymbol{r}, \alpha)=R_{n l}\left(Z_{l m}(\alpha), r\right) Y_{l m}(\vartheta, \varphi)
\ee
with eigenenergies
\be
E_{(n l) m}^{\rm 3D,aniso}(\alpha)=-\frac{1}{n^2} Z_{l|m|}^2(\alpha),
\ee
where 
\be
Z_{l m}(\alpha)=\int_{\Omega} \mathrm{d} \Omega \frac{\left[Y_{l m}(\Omega)\right]^2}{\left[1-\alpha \cos ^2 \vartheta\right]^{1 / 2}} 
\ee
are the effective charges.

While $n$ and $l$ are only approximate quantum numbers, since $\tilde{H}$ in Eq.~(\ref{eq:exciton_GaSe}) does not commute with the total angular momentum $l$ but only commutes with the angular momentum in $z$-direction $l_z$ and the parity operator $\Pi$, only the magnetic quantum number $m$ and the parity (even or odd) remain good quantum numbers. Thus, we introduce the combined quantum number $m_\pm$ with $m_+$ for even parity and $m_-$ for odd parity. To provide more insight, the exact eigenfunctions of $\tilde{H}$ are also eigenfunctions of $l_z$ and $\Pi$. This means that while the $m$ states do not mix, the even and odd eigenfunctions are superpositions of $l=2\nu$ and $l=2\nu+1$ states, respectively, with $\nu$ being 0 or a positive integer. For example, considering only states up to $n=3$, the anisotropic exciton eigenstates are superpositions of
\begin{itemize}
\item 1$s$, 2$s$, 3$s$, and 3$d_{m_+}$ states with $m_+=0$ (even),
\item 2$p_{m_-}$ and 3$p_{m_-}$ states with $m_-=0$ (odd),
\item 2$p_{m_-}$ and 3$p_{m_-}$ states with $m_-=+1$ (odd), 
\item 2$p_{m_-}$ and 3$p_{m_-}$ states with $m_-=-1$ (odd), 
\item 3$d_{m_+}$ state with $m_+=+1$ (even), 
\item 3$d_{m_+}$  state with $m_+=-1$ (even),
\item 3$d_{m_+}$ state with $m_+=+2$ (even),
\item 3$d_{m_+}$ state with $m_+=-2$ (even).
\end{itemize}
According to Ref.~\cite{Deverin1969}, the ground state of the exciton for $\gamma=2$ consists approximately of the doublet $\left|3d_{m_+=\pm 2}\right>$.

For ML TMCs, we can consider the 2D limit ($\gamma=0$, i.e. $\mu_{\|}\rightarrow\infty$), in which the exciton binding energies are determined by the 2D hydrogen model, i.e. \cite{Vasko1999}
\be
E_{nm}^{\rm 2D}=-\frac{e^4\mu_{\perp}}{2\varepsilon_{\perp}^2\hbar^2}\frac{1}{(n+|m|+1/2)^2},
\ee
where $n=1,2,\ldots$ is the radial quantum number. The allowed magnetic quantum numbers are $\left|m\right|=0,1,2,\ldots,n-1$. This formula would suggest that as $|m|$ increases, the exciton binding energy decreases. However, while in 3D insulators and undoped semiconductors the macroscopic screening is well described by static dielectric constants, in 2D insulators and undoped semiconductors the macroscopic screening is nonlocal with a logarithmic divergence at small distances \cite{Keldysh1979,Cudazzo2011,Trolle&Pedersen2017}. At large distances the unscreened Coulomb interaction is recovered. The main consequence is that excitons with same $n$ but larger value of $m$ have smaller eigenenergies  $E_{nm}<E_{nm'}$ if $m>m'$, i.e. larger exciton binding energies. This unusual exciton behavior is called the dielectric confinement effect and has been discovered by the experimental observation of dark excitons in WS$_2$ and by comparison with GW-BSE calculations \cite{Ye2014}.

\subsection{Anisotropy Hamiltonian at the $\Gamma$-point}
Let us describe the first four valence bands by means of an minimal Hamiltonian that captures the effects of the crystal field and spin-orbit coupling (SOC).
For that, we start with a 8x8 \(\mathbf{k} \cdot \mathbf{p}\) Hamiltonian for GaSe, including both s and p orbitals, spin-orbit coupling, and uniaxial anisotropy, which is given by
\be
H = H_{\text{kin}} + H_{\text{CF}} + H_{\text{SOC}},
\label{eq:H_aniso}
\ee
The crystal field Hamiltonian $H_{\text{CF}}$ for uniaxial anisotropy can be expressed as
\be
H_{\text{CF}} = \Delta_\perp \left| p_x \right\rangle \left\langle p_x \right| + \Delta_\perp \left| p_y \right\rangle \left\langle p_y \right| + \Delta_\parallel \left| p_z \right\rangle \left\langle p_z \right|,
\ee
where the crystal fields are $\Delta_\perp$ and $\Delta_\parallel$.
The SOC Hamiltonian \(H_{\text{SOC}}\) can be written in terms of the orbital angular momentum operator \(\mathbf{L}\) and the spin angular momentum operator \(\mathbf{S}\):
\be
H_{\text{SOC}} = \lambda \mathbf{L} \cdot \mathbf{S}=\lambda \left[ L_z S_z + \frac{1}{2} (L_+ S_- + L_- S_+) \right].
\ee
Since the Hamiltonian with respect to the s-orbital states is block-diagonal and diagonal with respect to $H_{\text{CF}} + H_{\text{SOC}}$, we can omit it from further analysis. In the basis $\left\{\left|p_0, \uparrow\right>, \left|p_0, \downarrow\right>, \left|p_{+1}, \uparrow\right>, \left|p_{+1}, \downarrow\right>,  \left|p_{-1}, \uparrow\right>, \left|p_{-1}, \downarrow\right>\right\}$, the non-zero components of the spin angular momentum operator are
\begin{eqnarray}
\langle p_{+1}, \uparrow | L_z S_z | p_{+1}, \uparrow \rangle & = & \langle p_{-1}, \downarrow | L_z S_z | p_{-1}, \downarrow \rangle = \frac{\hbar^2}{2}, \nn\\
\langle p_{+1}, \downarrow | L_z S_z | p_{+1}, \downarrow \rangle & = & \langle p_{-1}, \uparrow | L_z S_z | p_{-1}, \uparrow \rangle = -\frac{\hbar^2}{2}, \nn\\
\langle p_{+1}, \downarrow | L_+ S_- | p_0, \uparrow \rangle & = &
\langle p_0, \downarrow | L_+ S_- | p_{-1}, \uparrow \rangle = \hbar^2 \sqrt{2}, \nn\\
\langle p_0, \uparrow | L_- S_+ | p_{+1}, \downarrow \rangle & = &
\langle p_{-1}, \uparrow | L_- S_+ | p_0, \downarrow \rangle = \hbar^2 \sqrt{2}.
\end{eqnarray} 
Thus, the SOC Hamiltonian matrix is
\be
H_{\text{SOC}} = \hbar^2\lambda \begin{pmatrix}
0 & 0 & 0 & \sqrt{2} & 0 & 0 \\
0 & 0 & 0 & 0 & \sqrt{2} & 0 \\
0 & 0 & \frac{1}{2} & 0 & 0 & 0 \\
\sqrt{2} & 0 & 0 & -\frac{1}{2} & 0 & 0 \\
0 & \sqrt{2} & 0 & 0 & -\frac{1}{2} & 0 \\
0 & 0 & 0 & 0 & 0 & \frac{1}{2}
\end{pmatrix}
\ee
The combined crystal field and SOC Hamiltonian matrix is then
\be
H_{\text{CF+SOC}} =  \begin{pmatrix}
\Delta & 0 & 0 & \hbar^2\lambda \sqrt{2} & 0 & 0 \\
0 & \Delta & 0 & 0 & \hbar^2\lambda \sqrt{2} & 0 \\
0 & 0 & \frac{\hbar^2\lambda}{2} & 0 & 0 & 0 \\
\hbar^2\lambda \sqrt{2} & 0 & 0 & -\frac{\hbar^2\lambda}{2} & 0 & 0 \\
0 & \hbar^2\lambda \sqrt{2} & 0 & 0 & -\frac{\hbar^2\lambda}{2} & 0 \\
0 & 0 & 0 & 0 & 0 & \frac{\hbar^2\lambda}{2}
\end{pmatrix},
\ee
where $\Delta=\Delta_\perp-\Delta_\parallel$. In the case that $\Delta\gg \lambda$, the crystal field-SOC Hamiltonian matrix is approximately diagonal, i.e.
\be
H_{\text{CF+SOC}} \approx  \begin{pmatrix}
\Delta & 0 & 0 & 0 & 0 & 0 \\
0 & \Delta & 0 & 0 & 0 & 0 \\
0 & 0 & \frac{\hbar^2\lambda}{2} & 0 & 0 & 0 \\
0 & 0 & 0 & -\frac{\hbar^2\lambda}{2} & 0 & 0 \\
0 & 0 & 0 & 0 & -\frac{\hbar^2\lambda}{2} & 0 \\
0 & 0 & 0 & 0 & 0 & \frac{\hbar^2\lambda}{2}
\end{pmatrix}.
\label{eq:H_crystal-SOC}
\ee
Thus, the basis $\{\left|s, \uparrow\right>, \left|s, \downarrow\right>, \left|p_0, \uparrow\right>, \left|p_0, \downarrow\right>, \left|p_{+1}, \uparrow\right>, \left|p_{+1}, \downarrow\right>,$  $\left|p_{-1}, \uparrow\right>, \left|p_{-1}, \downarrow\right>\}$ is a good approximation for the eigenstates in the case $\Delta\gg \lambda$. Note that these eigenstates can be labeled by $\left|m_{j,\pm}\right>$, where $m_{j,\pm}$ is the magnetic quantum number of the total angular momentum $j$ and $\pm$ labels the parity. 

\subsection{Linear Response}
In the static approximation of the screening the effective two-particle Hamiltonian in Eq.~(\ref{H_2p}) can be simplified to
\be
\begin{gathered}
\sum_{n_3 n_4}\left\{\left(\epsilon_{n_2}-\epsilon_{n_1}\right) \delta_{n_1 n_3} \delta_{n_2 n_4}+\left(f_{n_1}-f_{n_2}\right)\left[\bar{v}_{\left(n_1 n_2\right)\left(n_3 n_4\right)}\right.\right. \\
-\left.\left.W_{\left(n_1 n_2\right)\left(n_3 n_4\right)}\right]\right\} A_\lambda^{\left(n_3 n_4\right)}=E_\lambda A_\lambda^{\left(n_1 n_2\right)}
\end{gathered}
\ee
considering only the resonant part of the effective two-particle Hamiltonian in Eq.~(\ref{H_2p}) it is possible to write the susceptibility in LR as
\be
\chi_{ij}^{(1)}(\omega) = \frac{N}{\epsilon_0}\sum_\lambda \frac{B^i_{g\lambda} B^j_{\lambda g}f_{g\lambda\bk}}{\hbar\omega-E_\lambda},
\label{eq:Chi1X}
\ee
where 
\begin{eqnarray}
B^j_{\lambda g} & = &  \left<\Psi_{\lambda}\left|\mu^j\right|g\right> 
= \sum_{n n'\bk}\mu_{n\bk,n'\bk}^j A_{\lambda}^{n n'} \nn\\
& = & \frac{1}{\sqrt{N}}\sum_{n n'\bk}\mu_{n\bk,n'\bk}^j
\int d\mathbf{r}\phi_{nn'\lambda}(r)e^{i{\bf{k}}\cdot{\bf{r}}},
\label{eq:excitonB}
\end{eqnarray}
where $d$ is the dimension of integration. $f_{nm\bk}=f_{n\bk}-f_{m\bk}$ is the difference between two Fermi functions.
$\left|g\right>$ is the ground state of the semiconductor, and $\mu_{n,n'}^j$ is the $j$th component electric dipole matrix vector $\bmu$ in the length gauge,
and the dielectric function in LR is
\be
\varepsilon^{(1)}(\omega)=1-\lim_{\bq \rightarrow 0} v_0(\mathbf{q}) \chi^{(1)}(\omega).
\ee

\subsection{Going out of equilibrium with Kadanoff-Baym Approach}


The only approach to describe interaction of electronic system with strong electromagnetic irradiation by first-principles is the technique of nonequilibrium Green's functions (NEGF) \cite{Schafer2002,Kremp2004} with the help of dynamical KBEs \cite{Pines1994}. Out of equilibrium dynamics of the electronic states, which causes the NLR, is represented by KBEs, which couple equations for greater and lesser Green's functions, $G^>$ and $G^<$. Considering only long-wavelength perturbations, we can neglect off-diagonal elements $(\mathbf{k},\mathbf{k'})$, since momentum $\mathbf{k}$ is conserved.  Using Hedin's COHSEX approximation to the self-energy \cite{Hedin1999}, we can significantly simplify the KBEs, since in the static limit of the self-energy \cite{Schafer2002,Kremp2004,Spataru2004} we can neglect $(t,t')$ off-diagonal elements. In this approach, it is possible to extract from the dynamical KBEs for $G^>$ and $G^<$ a closed equation for $G^<$ only \cite{Attaccalite2011}. The KBE in terms of the lesser Green's function matrix in band space is then
\begin{equation}
\label{KBE}
\begin{split}
i\hbar {\partial  \over {\partial t}} \mathbf{G}^<_{\bk,nn'}\left( t \right )   =  \left[ {\mathbf{H}}_{\bf{k}}, \mathbf{G}_{\mathbf{k}}^{<} \left( t \right) \right]_{nn'}.
\end{split}
\end{equation}
The Hamiltonian matrix in band space is given by
\begin{equation}
{\mathbf{H}}_{\bf{k}}={\mathbf{h}}_{\bf{k}}^{GW} + {\Delta\mathbf{V}}_{\bf{k}}^{H} + \Delta\Sigma_{\mathbf{k}}^{\textnormal{COHSEX}} + {\mathbf{U}}_{\bf{k}},
\end{equation}
with matrix elements $H_{nn'}(\bk)$. Here $\mathbf{h}_{\bf{k}}^{GW} = \mathbf{h}_{\bf{k}} + \Delta \mathbf{h}_{\bf{k}}$, 
$\mathbf{h}_{\bf{k}}$ is KS Hamiltonian, and ${\Delta\mathbf{h}}_{\bf{k}}$ is the $GW$ correction to the KS eigenvalues. The bold notation in Eq.~\eqref{KBE} with index $\mathbf{k}$ means matrix elements between the two KS eigenstates  
\be
\mathbf{G}_{\bk,nn'}^{<}\left( t \right) = \left< \phi_{n\bk} \left( \mathbf{r}_1 \right) | G^{<} \left( \mathbf{r}_1,\mathbf{r}_2, t \right) | \phi_{n'\bk} \left( \mathbf{r}_2 \right) \right>
\ee
${\Delta\mathbf{V}}_{\bf{k}}^{H}$ and $\Delta\Sigma_{\mathbf{k}}^{\textnormal{COHSEX}}$ are the difference between perturbed and unperturbed electronic systems for Hartree potential and COHSEX self-energy,
\be
{\Delta\mathbf{V}}_{\bf{k}}^{H} = {\mathbf{V}}_{\bf{k}}^{H}[\rho] - {\mathbf{V}}_{\bf{k}}^{H}[\tilde{\rho]}$$
$$\Delta\Sigma_{\mathbf{k}}^{\textnormal{COHSEX}} = \Sigma_{\mathbf{k}}^{\textnormal{COHSEX}}\left[ G^< \right] - \Sigma_{\mathbf{k}}^{\textnormal{COHSEX}}\left[ \tilde{G}^< \right].
\ee
${\mathbf{U}}_{\bf{k}} = < \phi_{n\mathbf{k}} | U | \phi_{n'\mathbf{k}} >$ represents the electron-light interaction, and in the length gauge the interaction potential is
\begin{equation}
U = -e\mathbf{r}\cdot \mathbf{E} (t)
\end{equation}
The electromagnetic radiation is switched on at $t=0$, and the out-of-equilibrium dynamics of the electronic configuration is evolving in time according to the KBE \eqref{KBE} in terms of the lesser Green's function $\mathbf{G}_\mathbf{k}^{<}\left( t \right)$.

When the external perturbation $U$ is weak, the KBE \eqref{KBE} is equivalent to the BSE \eqref{BSE_L} \cite{Attaccalite2011}.  
To show that, it is necessary to take the functional derivative of Eq.~\eqref{KBE} with respect to the perturbation $U$ and to introduce the retarded density-density correlation function
\begin{equation}
\label{correlation}
\chi^r(\mathbf{r},t;\mathbf{r}^{\prime},t^{\prime})  = \frac{\left<\delta\rho(\mathbf{r}t)\right>}{\delta U(\mathbf{r}^{\prime}t^{\prime})} \biggr|_{U=0}.
\end{equation}
Then using definition of charge density from the Green's function $G^{<}$ as
\be
\rho \left( \mathbf{r},t \right) = \frac{i}{\hbar} \sum_{{n}{n'}\mathbf{k}} \phi_{{n}{\mathbf{k}}} \left( \mathbf{r} \right) \phi^{*}_{{n'}{\mathbf{k}}} \left( \mathbf{r} \right) \mathbf{G}^{<}_{{\mathbf{k}}} \left( t \right), 
\ee
considering only optical response ($\mathbf{q}=0$), and calculating the matrix elements of \eqref{correlation}, one can obtain
\begin{equation}
\label{correlation2}
\chi^r_{nn',\mathbf{k},lm,\mathbf{p}}(t,t^{\prime})  = \frac{\delta \left<iG^{<}_{nn',\bk}(t)\right>}{\delta U_{lm,\mathbf{p}}(t^{\prime})} \biggr|_{U=0}
\end{equation}
Then keeping only linear terms with respect to $U$, the equivalence between the KBE in Eq.~\eqref{KBE} and the BSE in Eq.~\eqref{BSE_ChiG0W} for $\chi^r$ can be shown.

\subsection{Dynamical Polarization and Nonlinear Response}
When the perturbation by the electromagnetic wave $\mathbf{E}(t)$ is strong enough to produce NLR, the polarization $\mathbf{P}$ of the electronic system can be expanded in a power series of $\mathbf{E}$, which takes the following form in the frequency domain,
\begin{equation}
\begin{split}
&\mathbf{P} = {\mathbf{P}_0} + \\
&+ {\chi ^{\left( 1 \right)}}\left( \omega _1 \right)\mathbf{E}_1\left( \omega _1  \right) +  \\
&+ {\chi ^{\left( 2 \right)}}\left( {{\omega _1}, {\omega _2}} \right)\mathbf{E}_1\left( {{\omega _1}} \right)\mathbf{E}_2\left( {{\omega _2}} \right) +   \\
&+ {\chi ^{\left( 3 \right)}}\left( {{\omega _1}, {\omega _2}, {\omega _3}} \right)\mathbf{E}_1\left( {{\omega _1}} \right)\mathbf{E}_2\left( {{\omega _2}} \right)\mathbf{E}_3\left( {{\omega _3}} \right) + ... 
\end{split}
\label{eq:Polar}
\end{equation} 
Here $\mathbf{P}_0$ is the intrinsic static polarization of the material (e.g. ferroelectricity), $\chi^{(i)}$  is the $i$-th order susceptibility tensors, and $\mathbf{E}(t)$ is the sum of electromagnetic waves with different frequencies $\omega _1$, $\omega _2$, and $\omega _3$, affecting the electronic system, i.e.
\begin{equation}
\begin{split}
\mathbf{E}\left( t \right) = {{\mathbf{E}}^0_1}\sin \left( {{\omega _1}t} \right) + {\mathbf{E}}^0_2\sin \left( {{\omega _2}t} \right) + {{\mathbf{E}}^0_3}\sin \left( {{\omega _3}t} \right) + ...
\end{split}
\end{equation} 
The polarization $\mathbf{P}$ for periodic systems is an ill-defined quantity and depends on how one selects the unit cell. King-Smith and Vanderbilt \cite{KingSmith1993} were able to uniquely define the polarization for crystals with the help of Berry's phase \cite{Resta1994,Resta1998}, which corresponds to the phase difference of wavefunctions between two ground states $\xi_1$ and $\xi_2$,
\begin{equation}
{{e^{-i\Delta_{12}}} } = \frac{ \langle \Psi {\left( \xi_1 \right)} | \Psi {\left( \xi_2 \right)}}{ | \langle \Psi {\left( \xi_1 \right)} | \Psi {\left( \xi_2 \right)} |}  
\end{equation}
or
\begin{equation}
{{\Delta_{12}} } = -{\mathop{\rm Im}\nolimits} \ln \left\langle {{\Psi\left(\xi_1 \right) }\left|  {{\Psi \left(\xi_2 \right)}} \right.} \right\rangle .
\end{equation}
For isolated systems, the polarization is defined through the electric dipole operator $\bmu_\alpha=e\br_{\alpha}$ as
\begin{equation}
{{\mathbf{P}}_\alpha } \sim \int \mathbf{r_{\alpha}} \rho \left( \mathbf{r} \right) d\mathbf{r} = \sum\limits_{n} \langle \phi_n | \mathbf{r_{\alpha}} | \phi_n \rangle,
\end{equation}
where $| \phi_n \rangle$ are the eigenstates of a finite system.
In contrast to that, for periodic systems the electric dipole operator is replaced by $-ei\hbar\bnabla_\bk$, leading to
\begin{equation}
{{\mathbf{P}}_\alpha } \sim \int d{\mathbf{k}} \sum\limits_{n} \left< u_{n\bk}\left| i\hbar\bnabla_{k_{\alpha}} \right| u_{n\bk}\right> ,
\end{equation}
where $\left|u_{n\bk}\right>$ is the periodic part of the Bloch state $\left|\psi_{n\bk}\right>$. 

Note that the electric dipole matrix elements $\bmu_{nn'}=e \br_{nn'}$ are defined in the length gauge by \cite{Blount1962}
\be
\br_{nn'} = -i\hbar\left<\psi_{n\bk}\left|\bnabla_\bk\right|\psi_{n'\bk}\right>.
\ee
For a single-particle Hamiltonian of the form $H_0=\frac{p^2}{2m}+V(\br)-e\br\cdot\bE$, where $V(\br)$ is the periodic lattice potential with $V(\br)=V(\br+{\bf a})$, ${\bf a}$ being a crystal unit vector, it is possible to perform the transformation
\be
\tilde{H}_0(\bk)=e^{-i\bk\cdot\br}H_0e^{i\bk\cdot\br}=\frac{\left(\bp+\hbar\bk\right)^2}{2m}+V(\br)-e\br\cdot\bE.
\ee
The eigenstates of cell-periodic Hamiltonian $\tilde{H}_0(\bk)$ are the periodic parts of the Bloch states $\left|u_{n\bk}\right>$. Choosing this basis set is equivalent to using the Bloch state as an ansatz for the time-independent Schr\"odinger equation and obtaining the cell-periodic Schr\"odinger equation $\tilde{H}_0(\bk)\left|u_{n\bk}\right>=E_{n\bk}\left|u_{n\bk}\right>$.
Therefore, the bandstructure of a crystal can be expressed in terms of $\left|u_{n\bk}\right>$ with eigenenergies $E_{n\bk}$.
Thus, taking advantage of the cell-periodic Hamiltonian, it is possible to write the matrix elements of the position operator
\be
\br_{nn'}=-i\hbar \left<u_{n\bk}\left|\bnabla_\bk \right|u_{n'\bk}\right>=-\calA_{nn'}(\bk)
\ee
in terms of the Berry connections
\be
\calA_{nn'}(\bk) = i\hbar \left<u_{n\bk}\left|\bnabla_\bk \right|u_{n'\bk}\right>.
\ee
The corresponding Berry curvatures are given by
\be
\calB_{nn'}(\bk)=\bnabla_\bk\times \calA_{nn'}(\bk).
\ee
Then the integration over the whole Brillouin zone over $d\mathbf{k}$ gives the polarization \cite{KingSmith1993} 
\begin{equation}
\label{P}
{{\mathbf{P}}_\alpha } = {{e{N_{{{\mathbf{k}}_\alpha }}}{{\bf{a}}_\alpha }} \over {2\pi V}}{\mathop{\rm Im}\nolimits} \ln \left\langle {{\Psi _0}\left| {{e^{i{{\mathbf{q}}_\alpha } \cdot {\bf{\hat X}}}}} \right| {{\Psi _0}}} \right\rangle 
\end{equation}
Here $\mathbf{a}_{\alpha}$ are the primitive lattice vectors, $\alpha = 1, 2, 3$, ${\mathbf{\hat X}} = \sum\limits_{i = 1}^N {{{\mathbf{x}}_i}}$ is the expectation value of the electronic position for $N$-particle system, ${{\mathbf{q}}_\alpha } = {{{{\bf{b}}_\alpha }} \over {{N_{\bf{k}}}_\alpha }}$, where $\mathbf{b}_{\alpha}$ is the primitive reciprocal lattice vectors, $N_{\mathbf{k}_{\alpha}}$ is the number of $\mathbf{k}$-points along the $\alpha$ direction, and  $\mathbf{q}_{\alpha}$ is the smallest distance between two $\mathbf{k}$-points along the $\alpha$ direction.

\subsection{Linear and nonlinear optical selection rules for non-interacting electron-hole pairs}
\label{sec:selection_rules_eh_pairs}
The $\epsilon$-stacked GaS family of materials have $D_{3h}$ point group symmetry at the $\Gamma$-point \cite{Mooser1973,Tang2015}. 
The character table of $D_{3h}$ and its double group are shown in Tables~\ref{table_D_3h} and \ref{table_D_3h_DG} along with the irreducible representations (IRs) \cite{Koster}.

\begin{table}[h]%
\centering%
\begin{tabular}{ |c |c |c |c |c |c |c |c |c |}\hline
$D_{3h}$    &    $E$    &    $\sigma_{h}$    &    $2C_{3}$    &    $2S_{3}$    &    $3C_{2}$    &    $3\sigma_{v}$ & & \\ 
\hline
$A'_{1}=\Gamma_1$  & 1   & 1   & 1  & 1   &  1  &  1 &   & $x^2+y^2$, $z^2$ \\
 $A'_{2}=\Gamma_2$ & 1  & 1   & 1 &  1  & $-$1 & $-$1 & $S_z$ &  \\
 $E'=\Gamma_6$ & 2 & 2 & $-$1 & $-$1 & 0 &  0 & $\{x,y\}$ & $\{x^2-y^2,xy\}$ \\
 $A''_{1}=\Gamma_3$  & 1 & $-$1 & 1 &  $-$1 &  1 & $-$1 & & $zS_z$ \\
  $A''_{2}=\Gamma_4$ & 1 & $-$1 & 1 & $-$1 &  $-$1 &  1 & $z$ & \\
 $E''=\Gamma_5$ & 2 & $-$2 & $-$1 & 1 & 0 &  0 & $\{S_x,S_y\}$ & $\{xz,yz\}$ \\
\hline
 $D_{1/2}=\Gamma_7$  & 2    $-$2 & 0  & 1 $-$1 & $\sqrt{3}$  $-\sqrt{3}$ & 0 &  0 & $\phi_{\frac{1}{2},\pm \frac{1}{2}}$ & \\
 $2S_{1}=\Gamma_9$  & 2 $-$2 & 0 & $-$2  2 & 0  \hspace{0.3cm} 0 & 0 &  0 & $\phi_{\frac{3}{2},\pm \frac{3}{2}}$ & \\
 $2S_{2}=\Gamma_8$  & 2 $-$2 & 0 & 1  $-$1 & $-\sqrt{3}$ $\sqrt{3}$ & 0 &  0 & $\Gamma_7\times\Gamma_3$ &  \\
\hline\end{tabular}
\caption{Character table of the group $D_{3h}$. $E$, $\sigma_{h}$, $2C_{3}$, $2S_{3}$, $3C_{2}$, and $\sigma_{v}$ are the single group IRs and $D_{1/2}$, $2S_{1}$, $2S_{2}$ are the corresponding double group IRs.}
\label{table_D_3h}
\end{table}

\begin{table}[h]%
\centering%
\begin{tabular}{ |c |c |c |c |c |c |c |}\hline
 $\Gamma_{i}(D_{3h})$                          &    $A^{\prime}_{1}$    &    $A^{\prime}_{2}$    &    $A^{\prime\prime}_{1}$    &    $A^{\prime\prime}_{2}$    &    $E^{\prime}$    &    $E^{\prime\prime}$\\ 
\hline
$\Gamma_{i}\times D_{1/2}$ & $D_{1/2}$  & $D_{1/2}$  & $2S_2$                 & $2S_2$    &           $2S_1+2S_2$        &  $2S_1+D_{1/2}$\\
\hline 
\end{tabular}
\caption{Double-group representations obtained from single-group representation for $D_{3h}$ group.}
\label{table_D_3h_DG}
\end{table}

The conduction band (CB), valence band (VB), VB-1, VB-2, VB-3, and VB-4 states near the $\Gamma$-point consist mostly of Ga $s$-like, Se $p_z$-like, and Se $\{p_x,p_y\}$-like orbitals (see Supplementary Information) and
transform according to $A'_{1}=\Gamma_4$, $A''_{2}=\Gamma_1$, and $E'$ IRs of the $D_{3h}$ group, respectively. The VB is split from the \{VB-1,VB-3\} doublet and the \{VB-2,VB-4\} doublet states by the crystal-field anisotropy and SOC.
At the DFT-PBE level of accuracy without SOC, we obtain (see Fig.~\ref{fig:GaSe_unit_cell})
\bea
E_{\rm VB}-E_{\rm VB-1/VB-3} & = & 0.79 \text{ eV}, \nn\\
E_{\rm VB}-E_{\rm VB-2/VB-4} & = & 1.05 \text{ eV},
\eea
for bulk GaSe. Without SOC, the degeneracy between VB-1 and VB-3 and the degeneracy between VB-2 and VB-4 is due to the mixing of two sets of $\{p_x,p_y\}$-like orbitals of the two Se atoms in the unit cell.

Including SOC, we obtain (see Fig.~\ref{fig:GaSe_unit_cell})
\bea
E_{\rm VB}-E_{\rm VB-1} & = & 0.66 \text{ eV}, \nn\\
E_{\rm VB}-E_{\rm VB-2} & = & 0.92 \text{ eV}, \nn\\
E_{\rm VB}-E_{\rm VB-3} & = & 1.02 \text{ eV}, \nn\\
E_{\rm VB}-E_{\rm VB-4} & = & 1.22 \text{ eV}.
\eea
At the GW level of accuracy including SOC, we obtain (see Fig.~\ref{fig:GaSe_unit_cell})
\bea
E_{\rm VB}-E_{\rm VB-1} & = & 0.48 \text{ eV}, \nn\\
E_{\rm VB}-E_{\rm VB-2} & = & 0.90 \text{ eV}, \nn\\
E_{\rm VB}-E_{\rm VB-3} & = & 0.92 \text{ eV}, \nn\\
E_{\rm VB}-E_{\rm VB-4} & = & 1.25 \text{ eV}.
\eea
While spin-orbit splitting is introduced into the
$\{p_x,p_y\}$-like orbitals, Kramers degeneracy is present, as can be seen from the eigenenergies in Eq.~\eqref{eq:H_crystal-SOC}.

\begin{figure}
    \includegraphics{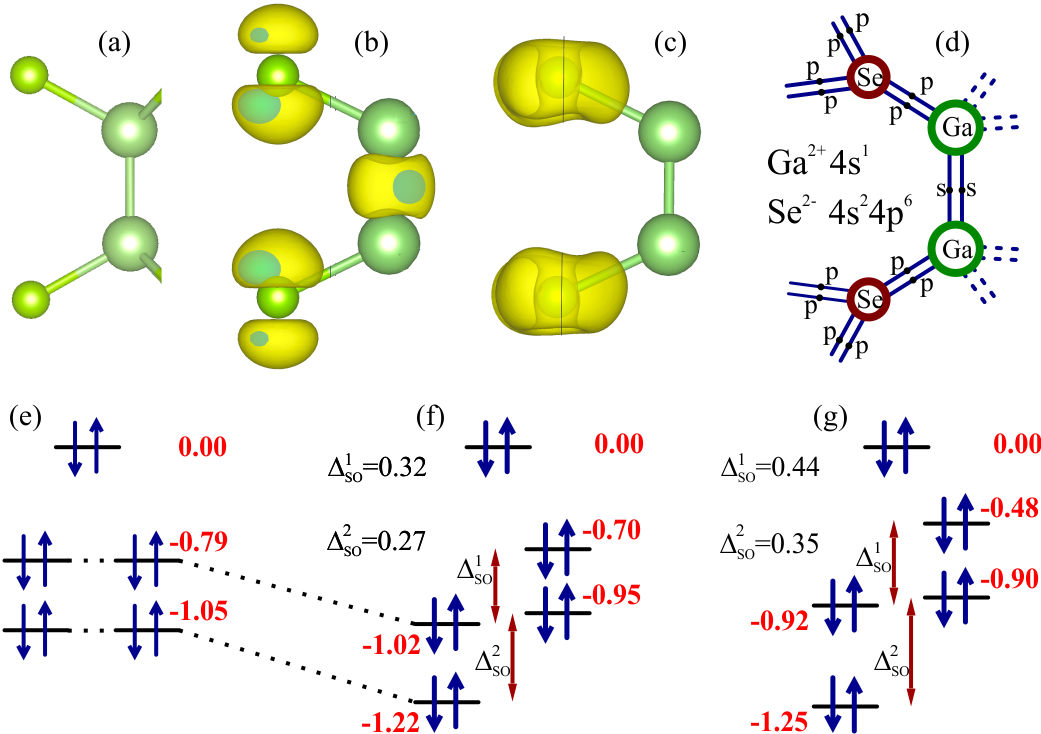}
    \caption{(a) Unit cell of 2D GaSe consisting of two Ga atoms (big green balls) and two Se atoms (small balls). (b)-(c) Electron charge density (shown by yellow) for (b) VB orbital and (c) VB-1,VB-2,VB-3, and VB-4 orbitals. (d) Bond formation in GaSe. (e)-(g) Energy levels at $\Gamma$ (in eV) for VB, VB-1, VB-2, VB-3, and VB-4 for (e) no SOC, (f) including SOC, and (g) SOC-$GW$.}
\label{fig:GaSe_unit_cell}
\end{figure}

Let us first analyze the optical selection rules for non-interacting states, i.e. in the independent-particle approximation (IPA).
The matrix elements of the dielectric tensor in LR in three dimensions ($i,j=x,y,z$) are determined by the Kubo–Greenwood formula \cite{Ivliev2017} for the electric susceptibility
\begin{equation}\label{eq:KGW}
\chi_{ij}(\omega)=\frac{n}{\epsilon_0\hbar}\sum_{nm\bf{k}}\frac{\mu_{nm}^{i}\mu_{mn}^{j}f_{nm\bk}}{[\omega_{nm}({\bf{k}})-\omega-i\Gamma/{\hbar}]}
\end{equation}  
where $\mu_{nm}^{j}=e\langle u_{n\bk}|r^{j}|u_{m\bk}\rangle$ is the electric dipole matrix, $n$ is the density of the crystal, $f_{nm\bk}=f_{n\bk}-f_{m\bk}$ is the difference between two Fermi functions, and $\Gamma=0.01$ eV the broadening. $\hbar\omega_{nm}(\bk)=E_{n\bk}-E_{m\bk}$ is the energy difference between the eigenenergies $E_{n\bk}$ and $E_{m\bk}$ of the Bloch states $\left|n\bk\right>$ and $\left|m\bk\right>$.

A general result from group theory states that an optical transition is allowed by symmetry only if the direct product of IRs $\Gamma(|u_{n\bk}\rangle)\otimes$ $\Gamma(\mu^j)$$\otimes\Gamma(|u_{m\bk}\rangle)$ contains $\Gamma(I)$ in its decomposition in terms of a direct sum. $\Gamma(I)$ denotes the IR for the identity, i.e., $A'_1$ for $D_{3h}$.
The in-plane and out-of-plane components of $p_{vu}^{j}$ must be considered individually because they transform according to different IRs of the point group. The resulting optical selection rules are shown in Fig.~\ref{fig:SR_linear_ehpairs}, which agree with the ones obtained from group theory shown in Table~\ref{table_D_3h_selection_rules}. These selection rules corroborate the difference between the in-plane and out-of-plane band gaps $E_{g||}$ and $E_{g\perp}$, respectively, which can be seen in the in-plane and out-of-plane susceptibilities Im$[\chi_{\parallel}](\omega)$ and Im$[\chi_{\perp}](\omega)$, respectively.

\begin{table}[h]
\begin{tabular}{|c |c |c |c |c |c |c|}\hline
$D_{3h}$ & $A_{1}^{\prime}$ & $A_{2}^{\prime}$ & $A_{1}^{\prime\prime}$ & $A_{2}^{\prime\prime}$ & $E^{\prime}$ & $E^{\prime\prime}$   \\
\hline
$A_{1}^{\prime}$ & & &    & $\pi$      &  $\sigma$  &   \\
\hline
 $A_{2}^{\prime}$  &  &     & $\pi$   &               & $\sigma$ &     \\
\hline
$A_{1}^{\prime\prime}$  &      & $\pi $  & & &     & $\sigma$      \\
\hline
$A_{2}^{\prime\prime}$ &  $\pi$ & & & & & $\sigma$\\
\hline
$E^{\prime}$ & $\sigma$&$\sigma$ & & & $\sigma$&$\pi$\\
\hline
$E^{\prime\prime}$ & & & $\sigma$&$\sigma$ & $\pi$&$\sigma$\\
\hline
\end{tabular}
\caption{Electric dipole selection rules for noninteracting electron-hole pairs in $D_{3h}$ symmetry. $\sigma$ represents in-plane transitions while $\pi$ represents out-of-plane transitions.}
\label{table_D_3h_selection_rules}
\end{table}

\begin{figure}[htb]
\begin{center}
\includegraphics[width=3.0in]{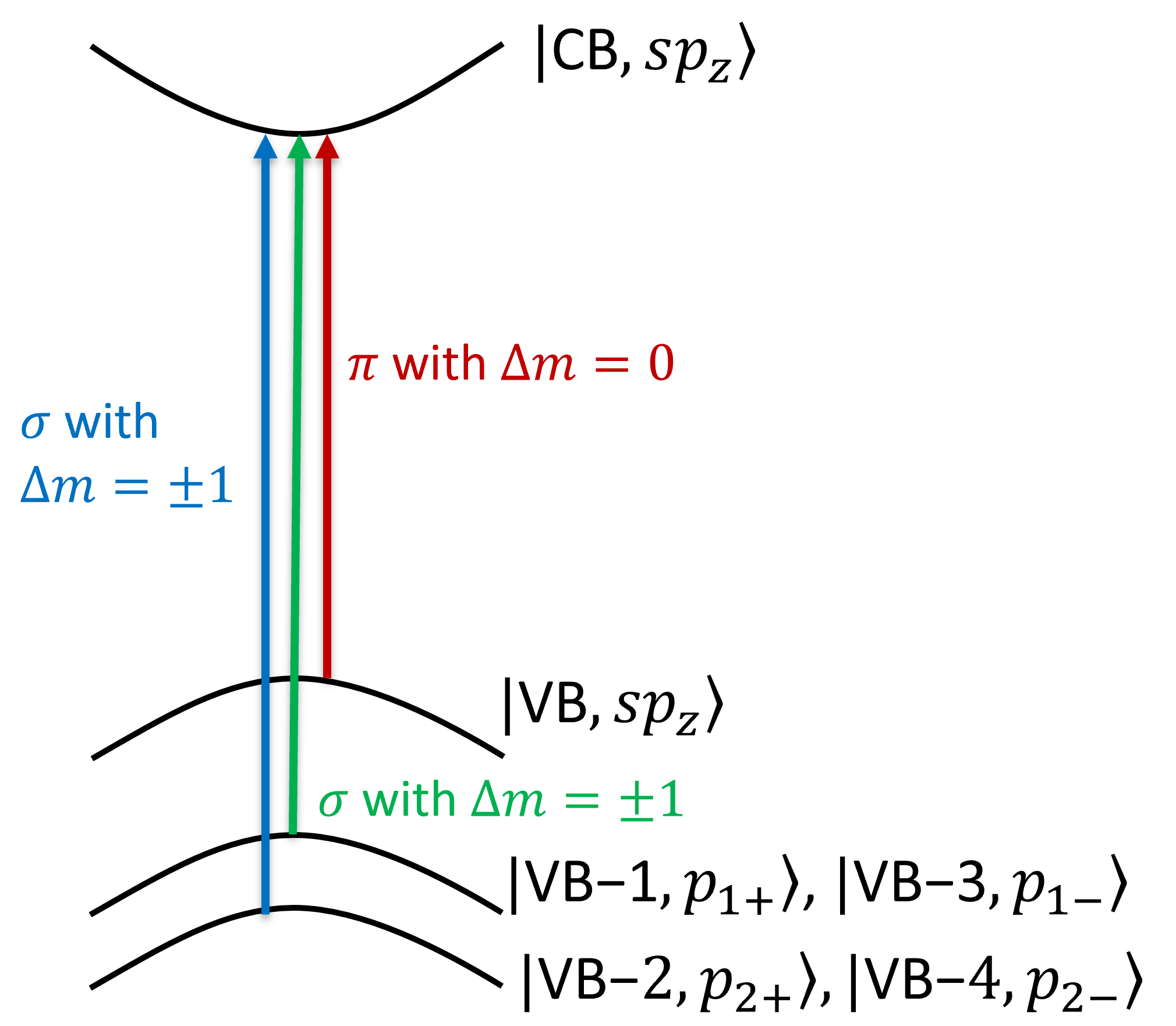}
\end{center}
\caption{Allowed optical transitions in LR in TMCs for noninteracting electron-hole pairs.}
\label{fig:SR_linear_ehpairs}
\end{figure} 

After identifying the CB with the $A'_{1}$ IR and the VB with the $A''_{2}$ IR in Table~\ref{table_D_3h_selection_rules}, we find that the optical transition from VB to CB is only allowed for out-of-plane polarization, i.e. $z$-polarization, of the electric field, which corresponds to a $\pi$-transition.
Conversely, by identifying the \{VB-1,VB-3\} doublet and the \{VB-2,VB-4\} doublet with the $E'$ IR, we find that the optical transitions from the \{VB-1,VB-3\} doublet and the \{VB-2,VB-4\} doublet to the CB are allowed only for in-plane polarization, i.e. either $x$- or $y$-polarization, of the electric field, which corresponds to $\sigma$ transitions. The allowed optical transition in LR for noninteracting electron-hole pairs are shown in the level diagram in Fig.~\ref{fig:SR_linear_ehpairs}. Although the orbital angular momentum is not conserved and therefore $l$ is not a good quantum number due to the uniaxial anisotropy, the orbital angular momentum in $z$-direction is conserved, and therefore the magnetic quantum number $m$ is still a good quantum number in the case of uniaxial anisotropy. Consequently, the eigenstates can be labeled by the magnetic quantum number $m$, i.e. $\left|sp_z,m=0\right>$ and $\left|p_{\pm 1},m=\pm 1\right>$. This labeling clarifies the optical selection rules in terms of the conservation of the orbital angular momentum projected along the $z$-axis.

Turning to the second-order NLR, the susceptibility in second-order perturbation theory is given by \cite{Boyd_NLO}
\be
\chi_{i j k}^{(2)}\left(\omega_\sigma, \omega_q, \omega_p\right)=\frac{n}{\epsilon_0 \hbar^2} \mathcal{P}_F \sum_{l m n} \frac{\mu_{l n}^i \mu_{n m}^j \mu_{m l}^k f_{lm\bk}}{\left(\omega_{n l}-\omega_\sigma\right)\left(\omega_{m l}-\omega_p\right)},
\label{eq:chi2}
\ee
where we suppressed the line broadening parameters, and $\omega_\sigma=\omega_p+\omega_q$. The symbol $\mathcal{P}_F$ denotes the full permutation operator for the summation over all permutations of the frequencies $\omega_p, \omega_q$, and $\omega_\sigma$. Note that intraband electric dipole matrix elements can be nonzero according to the optical selection rules in Table~\ref{table_D_3h_selection_rules} for hybridized states, such as $sp$-hybridized states.

\begin{figure}[htb]
\begin{center}
\includegraphics[width=3.0in]{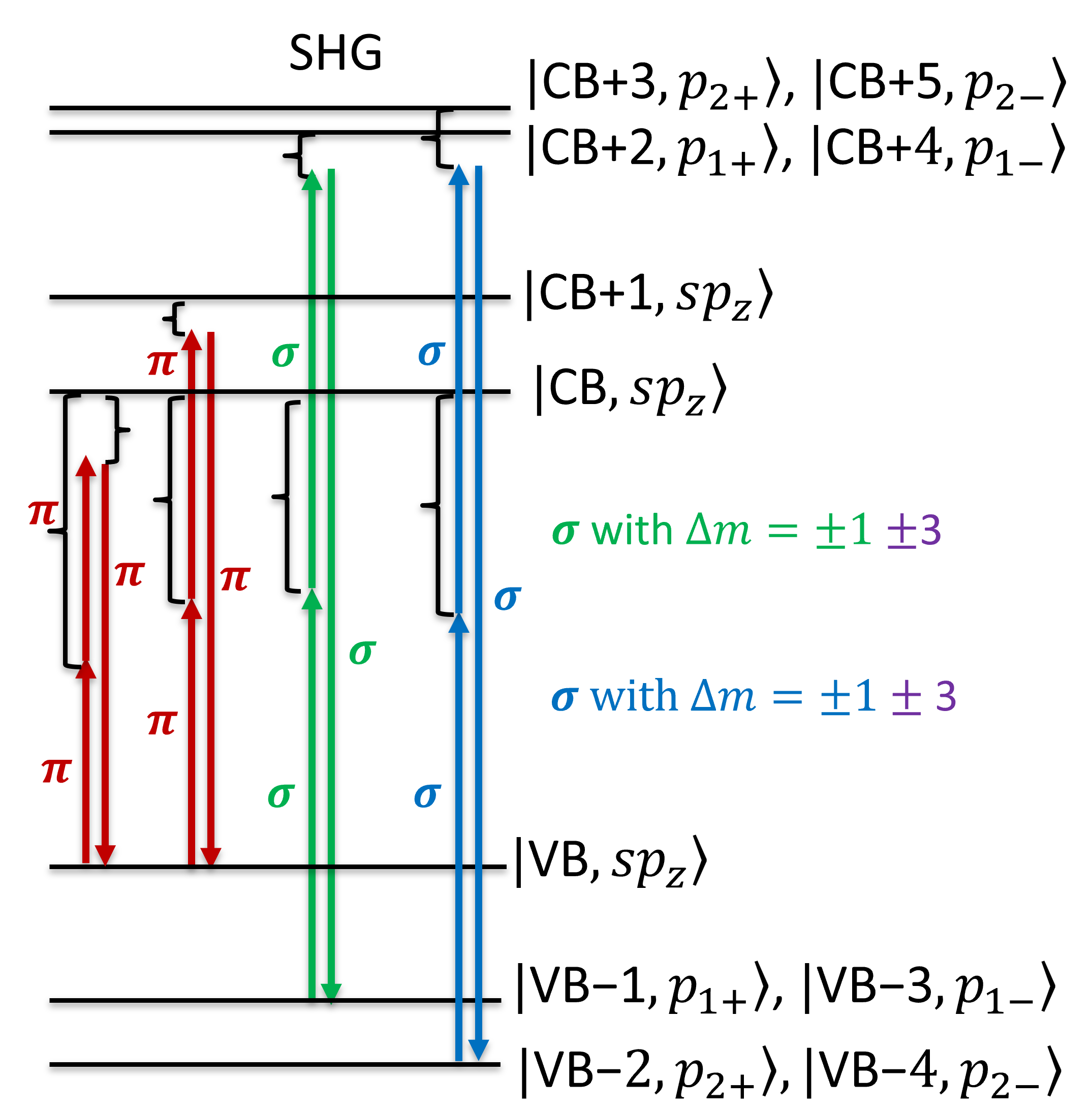}
\end{center}
\caption{Allowed optical transition responsible for SHG in NLR for noninteracting electron-hole pairs in TMCs driven by $\pi$-polarized, $\sigma^+$-polarized, and $\sigma^-$-polarized lasers. The curly brackets depict the detuning energies of the laser energy with respect to the resonant transition energies.}
\label{fig:SR_nonlinear_ehpairs_SHG}
\end{figure} 

In the case of SHG, the energy of the photons needs to be
below the two-photon absorption (2PA) energies. 
Then, the optical transitions are off-resonant out-of-plane ($z$ direction) driven by $\pi$-polarized photons ($\Delta m=0$) with main contribution from the paths 
\bea
\left|{\rm VB},sp_z^{m=0}\right>\rightarrow\left|{\rm CB},sp_z^{m=0}\right>\rightarrow\left|{\rm CB},sp_z^{m=0}\right>
\nn\\
\rightarrow\left|{\rm VB},sp_z^{m=0}\right>
\eea
and 
\bea
\left|{\rm VB},sp_z^{m=0}\right>\rightarrow\left|{\rm CB},sp_z^{m=0}\right>\rightarrow\left|{\rm CB}+1,sp_z^{m=0}\right> \nn\\
\rightarrow\left|{\rm VB},sp_z^{m=0}\right>
\eea
(see Fig.~\ref{fig:SR_nonlinear_ehpairs_SHG}). The off-resonant in-plane optical transitions are driven by virtual absorption of two $\sigma^-$ photons ($\Delta m=-2$) and the virtual emission of one $\sigma^+$ photon ($\Delta m=-(+1)$) or by virtual absorption of two $\sigma^+$ photons ($\Delta m=+2$) and the virtual emission of one $\sigma^-$ photon ($\Delta m=-(-1)$), with main contribution from the paths 
\bea
\left|\{{\rm VB}-3,{\rm VB}-4\},p_-^{m=-1}\right>\rightarrow\left|{\rm CB},sp_z^{m=0}\right>
\nn\\
\rightarrow\left|\{{\rm CB}+2,{\rm CB}+3\},p_+^{m=+1}\right> 
\nn\\
\rightarrow\left|\{{\rm VB}-3,{\rm VB}-4\},p_-^{m=-1}\right>
\eea
and 
\bea
\left|\{{\rm VB}-1,{\rm VB}-2\},p_+^{m=+1}\right>\rightarrow\left|{\rm CB},sp_z^{m=0}\right>
\nn\\
\rightarrow\left|\{{\rm CB}+4,{\rm CB}+5\},p_-^{m=-1}\right> 
\nn\\
\rightarrow\left|\{{\rm VB}-1,{\rm VB}-2\},p_+^{m=+1}\right>,
\eea
respectively. Note that the $\sigma$ transitions here are allowed due to the fact that the orbital angular momentum in z direction is conserved up to $\pm 3$ due to the $D_{3h}$ symmetry, i.e. $\Delta m=-1-1-(+1)+3=0$ and $\Delta m=+1+1-(-1)-3=0$, respectively. This off-resonant optical transition is described by the second-order susceptibility in Eq.~(\ref{eq:chi2}) and gives rise to SHG. The SHG level diagram with the allowed SHG transitions is shown for TMCs in Fig.~\ref{fig:SR_nonlinear_ehpairs_SHG}.

\begin{widetext}
In the case of third-order NLR, the susceptibility in third-order perturbation theory is given by \cite{Boyd_NLO}
\be
\chi_{k j i h}^{(3)}\left(\omega_\sigma, \omega_r, \omega_q, \omega_p\right)=\frac{n}{\epsilon_0 \hbar^3} \mathcal{P}_F \sum_{l m n \nu} \frac{\mu_{l \nu}^k \mu_{\nu n}^j \mu_{n m}^i \mu_{m l}^h f_{lm\bk}}{\left(\omega_{\nu l}-\omega_\sigma\right)\left(\omega_{n l}-\omega_q-\omega_p\right)\left(\omega_{m l}-\omega_p\right)},
\label{eq:chi3}
\ee
where $\omega_\sigma=\omega_p+\omega_q+\omega_r$. The third harmonic generation (THG) level diagram with allowed THG transitions for TMCs is shown in Fig.~\ref{fig:SR_nonlinear_ehpairs_THG}.
\end{widetext}

In the case of the 2PA process, two photons get absorbed resonantly. The main contribution for out-of-plane $\pi$ transition comes from the paths $\left|{\rm VB},sp_z,m=0\right>\rightarrow\left|{\rm CB},s,m=0\right>\rightarrow\left|{\rm CB},sp_z,m=0\right>$ and $\left|{\rm VB},sp_z,m=0\right>\rightarrow\left|{\rm CB},s,m=0\right>\rightarrow\left|{\rm CB}+1,sp_z,m=0\right>$, where the CB acts as a virtual (intermediate) state. In general, a 2PA is described by the third-order nonlinear susceptibility shown in Eq.~({\ref{eq:chi3}}). These 2PA resonance peaks are visible in Fig.~\ref{fig:X2X3_GaSe}. 
The main contribution for in-plane $\sigma$ transitions comes from the paths 
\bea
\left|\{{\rm VB}-3,{\rm VB}-4\},p_-^{m=-1}\right>\rightarrow\left|{\rm CB},sp_z^{m=0}\right>
\nn\\
\rightarrow\left|\{{\rm CB}+2,{\rm CB}+3\},p_+^{m=+1}\right>
\eea
and 
\bea
\left|\{{\rm VB}-1,{\rm VB}-2\},p_+^{m=+1}\right>\rightarrow\left|{\rm CB},sp_z^{m=0}\right>
\nn\\
\rightarrow\left|\{{\rm CB}+4,{\rm CB}+5\},p_-^{m=-1}\right>
\eea
for the resonant absorption of two $\sigma^+$ photons ($\Delta m=+2$)  or two $\sigma^-$ photons ($\Delta m=-2$), respectively.
The 2PA level diagram with the allowed 2PA transitions for TMCs is shown in Fig.~\ref{fig:SR_nonlinear_ehpairs_2PA}.

\begin{figure}[htb]
\begin{center}
\includegraphics[width=3.0in]{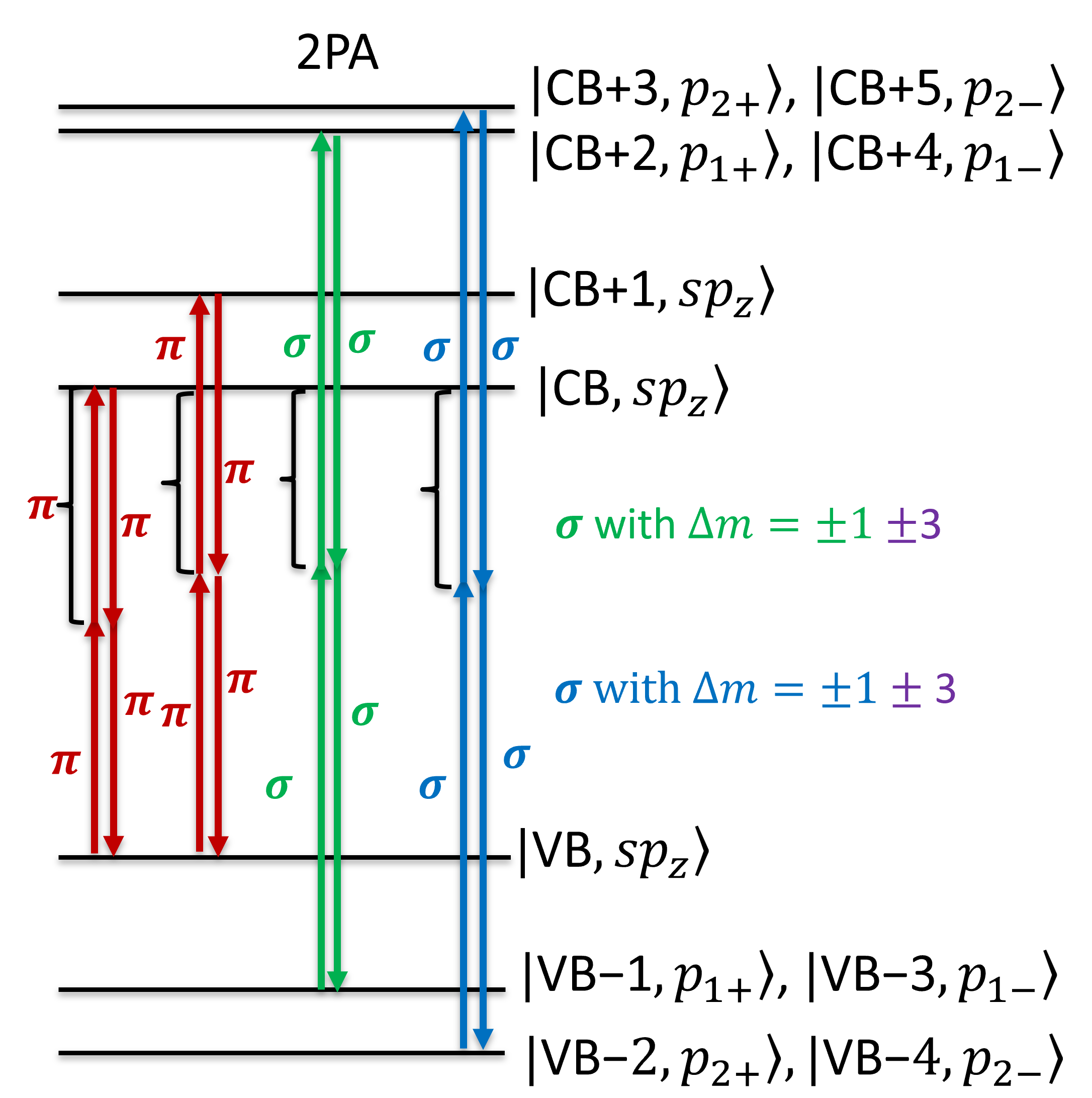}
\end{center}
\caption{Allowed optical transition responsible for 2PA in NLR for noninteracting electron-hole pairs in TMCs driven by $\pi$-polarized, $\sigma^+$-polarized, and $\sigma^-$-polarized lasers. The curly brackets depict the detuning energies of the laser energy with respect to the resonant transition energies. }
\label{fig:SR_nonlinear_ehpairs_2PA}
\end{figure} 

In the case of THG,
below the three-photon absorption (3PA) energies, the off-resonant out-of-plane (z direction) optical transition is driven by $\pi$ transitions ($\Delta m=0$) with main contribution from the paths 
\bea
\left|{\rm VB},sp_z^{m=0}\right>\rightarrow\left|{\rm CB},sp_z^{m=0}\right>
\rightarrow\left|{\rm CB},sp_z^{m=0}\right>
\nn\\
\rightarrow\left|{\rm CB},sp_z^{m=0}\right>
\rightarrow\left|{\rm VB},sp_z^{m=0}\right>
\eea
and 
\bea
\left|{\rm VB},sp_z^{m=0}\right>\rightarrow\left|{\rm CB},sp_z^{m=0}\right>
\rightarrow\left|{\rm CB},sp_z^{m=0}\right>
\nn\\
\rightarrow\left|{\rm CB}+1,sp_z^{m=0}\right>
\rightarrow\left|{\rm VB},sp_z^{m=0}\right>
\eea
(see Fig.~\ref{fig:SR_nonlinear_ehpairs_THG}). The off-resonant in-plane optical transitions are driven by the virtual absorption of three $\sigma$ photons and the virtual emission of one $\sigma$ photon ($\Delta m=0$ mod 3), with main contribution from the paths 
\bea
\left|\{{\rm VB}-1,{\rm VB}-2,{\rm VB}-3,{\rm VB}-4\},p_\pm^{m=\pm 1}\right>
\rightarrow\left|{\rm CB},sp_z^{m=0}\right>
\nn\\
\rightarrow\left|\{{\rm CB}+2,{\rm CB}+3,{\rm CB}+4,{\rm CB}+5\},p_\pm^{m=\pm 1}\right>
\nn\\
\rightarrow\left|\{{\rm CB}+2,{\rm CB}+3,{\rm CB}+4,{\rm CB}+5\},p_\pm^{m=\pm 1}\right> 
\nn\\
\rightarrow\left|\{{\rm VB}-1,{\rm VB}-2,{\rm VB}-3,{\rm VB}-4\},p_\pm^{m=\pm 1}\right>.
\eea
Note that the $\sigma$ transitions here are allowed due to the fact that the orbital angular momentum in z direction is conserved up to $\pm 3$ due to the $D_{3h}$ symmetry, i.e. $\Delta m=0$ mod 3. This off-resonant optical transition is described by the third-order susceptibility in Eq.~(\ref{eq:chi3}) and gives rise to THG. The THG level diagram with the allowed SHG transitions is shown for TMCs in Fig.~\ref{fig:SR_nonlinear_ehpairs_THG}.

\begin{figure}[htb]
\begin{center}
\includegraphics[width=3.0in]{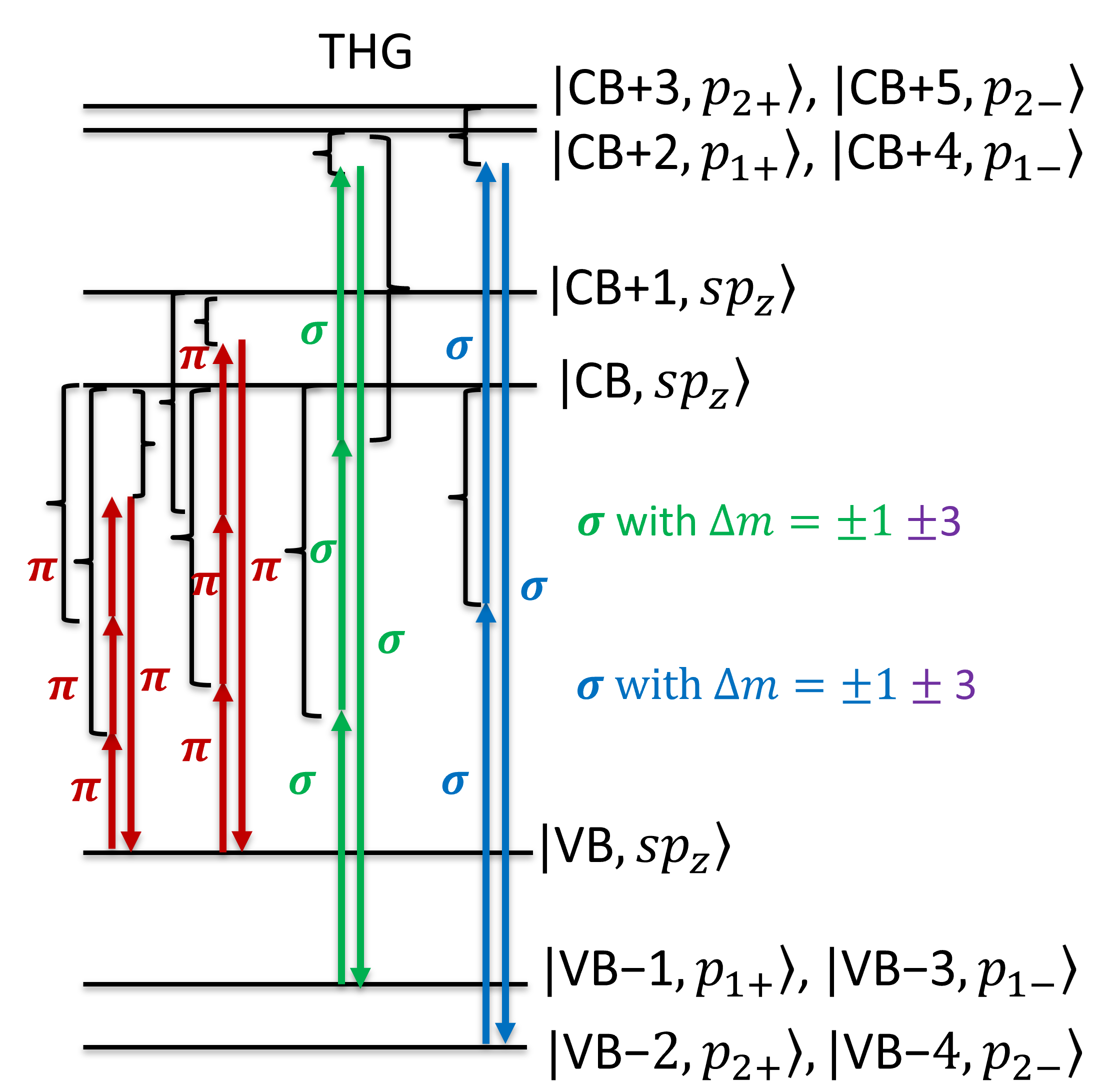}
\end{center}
\caption{Allowed optical transition responsible for THG in NLR for noninteracting electron-hole pairs in TMCs driven by $\pi$-polarized, $\sigma^+$-polarized, and $\sigma^-$-polarized lasers. The curly brackets depict the detuning energies of the laser energy with respect to the resonant transition energies.}
\label{fig:SR_nonlinear_ehpairs_THG}
\end{figure} 


\subsection{Linear and nonlinear optical selection rules for excitons}
\label{sec:selection_rules_excitons}
In the case of excitons we can generalize the above optical selection rules by including the relative motion of the electrons and holes. 
The linear susceptibility for excitons is shown in Eq.~({\ref{eq:Chi1X}}). 
The exciton matrix element with the ground state shown in Eq.~(\ref{eq:excitonB}) can be approximated by \cite{Yu&Cardona}
\begin{eqnarray}
B^j_{\lambda g} & = &  \frac{1}{\sqrt{N}}\sum_{n n'\bk}\mu_{n\bk,n'\bk}^j
\int d\mathbf{r}\phi_\lambda^{nn'}(r)e^{i{\bf{k}}\cdot{\bf{r}}} \nn\\
& \approx & \sqrt{N}\sum_{n n'}\mu_{n,n'}^j
\phi_\lambda^{nn'}(0),
\end{eqnarray}
if we assume that the electric dipole matrix element does not depend on $\bk$. In 3D for isotropic systems, this results in the optical selection rule $\lambda=1s,2s,3s,\ldots$, i.e. only $s$-wave ($l=0$) excitons can be excited in LR. A few examples of allowed optical transitions involving excitons are shown in Fig.~\ref{fig:SR_linear_excitons}. Compared with the linear selection rules for noninteracting electron-hole pairs, for excitons we need to consider also the symmetry of the envelope wavefunction. Therefore, group theory states for excitons that an optical transition in LR is allowed  by symmetry only if the direct product of IRs $\Gamma(\left|v\bf{k}\right>)\otimes \Gamma(\left|u\bf{k}\right>)\otimes \Gamma(\left|\phi_\lambda\right>)\otimes \Gamma(\mu^j)$ contains $\Gamma(I)$ in its decomposition in terms of a direct sum. 
As discussed above, if the dependence of the electric dipole moment on $\bk$ can be neglected, then the $s$ symmetry of the exciton envelope wavefunction does not change the optical selection rules imposed by symmetry when compared with the noninteracting electron-hole pairs.

\begin{figure}[htb]
\begin{center}
\includegraphics[width=3.0in]{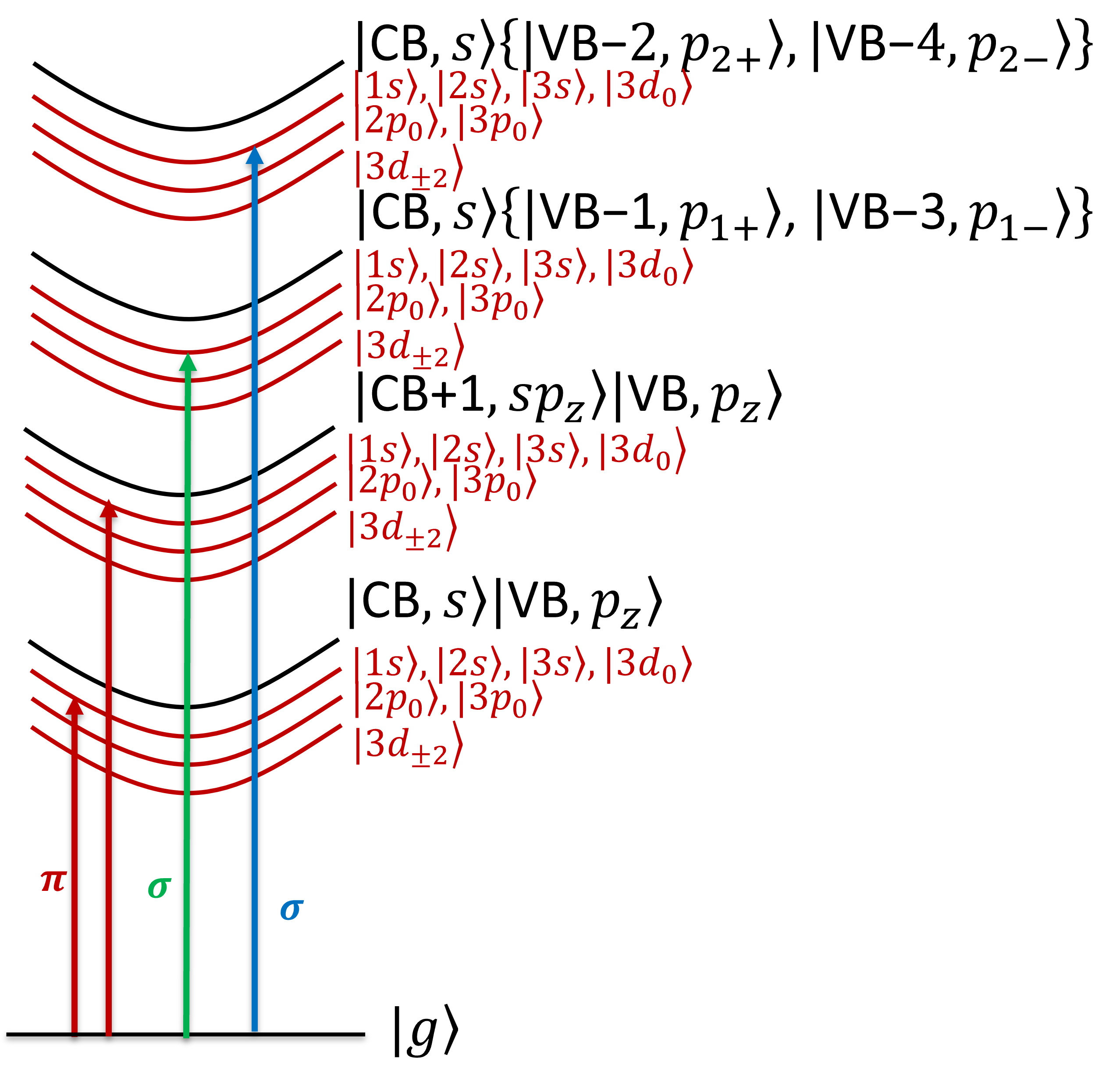}
\end{center}
\caption{Allowed optical transition in LR for excitons in TMCs shown in the two-particle diagram.}
\label{fig:SR_linear_excitons}
\end{figure} 

In the case of anisotropic excitons, according to our discussion in Sec.~\ref{sec:excitons_GaS}, we find that only eigenstates with $m_+=0$ can be excited in LR. Note that these anisotropic excitons can be more $s$-like or more $d$-like. In the case of bulk GaSe, perturbation theory shows that the states $\left|3s\right>$ and $\left|3d_0\right>$ are mixed \cite{Deverin1969}. The LR level diagram due to anisotropic excitons for TMCs is shown in Fig.~\ref{fig:SR_linear_excitons}. These optical selection rules are also valid in the 2D case.
Note that the spatially dependent screening ensures that excitons with the largest orbital angular momentum $|m|$ have the lowest energies, i.e. are the ground-state excitons. Consequently, according to the above optical selection rules, we expect the obtain dark excitons as ground-state excitons for ML TMCs using the GW-BSE approach, which is indeed the case, as shown in Sec.~\ref{sec:optical} below.

\begin{table}[h]
\begin{tabular}{|c |c |c |c |c |c |c|}\hline
$D_{3h}$ & $A_{1}^{\prime}$ & $A_{2}^{\prime}$ & $A_{1}^{\prime\prime}$ & $A_{2}^{\prime\prime}$ & $E^{\prime}$ & $E^{\prime\prime}$   \\
\hline
$A_{1}^{\prime}$ & & &    & $\pi$      &  $\sigma$  &   \\
\hline
 $A_{2}^{\prime}$  &  &     & $\pi$   &               & $\sigma$ &     \\
\hline
$A_{1}^{\prime\prime}$  &      & $\pi $  & & &     & $\sigma$      \\
\hline
$A_{2}^{\prime\prime}$ &  $\pi$ & & & & & $\sigma$\\
\hline
$E^{\prime}$ & $\sigma$&$\sigma$ & & & $\sigma$&$\pi$\\
\hline
$E^{\prime\prime}$ & & & $\sigma$&$\sigma$ & $\pi$&$\sigma$\\
\hline
\end{tabular}
\caption{Electric dipole selection rules for excitons in $D_{3h}$ symmetry. $\sigma$ represents in-plane transitions while $\pi$ represents out-of-plane transitions.}
\label{table_D_3h_selection_rules}
\end{table}

For NLR due to excitons, we need to calculate matrix elements of the electric dipole moment operator in terms of exciton eigenstates, i.e. 
\begin{eqnarray}
B^j_{\lambda_1 \lambda_2} & = & \left<\Psi_{\lambda_1}\left|\mu^j\right|\Psi_{\lambda_2}\right> \nn\\
& = & \sum_{n_1 n_1' n_2 n_2'\bk}\mu_{n_1\bk,n_2\bk}^j A_{\lambda_1}^{n_1 n_1'} A_{\lambda_2}^{n_2 n_2'}\delta_{n_1' n_2'} \nn\\
& & +\sum_{n_1 n_1' n_2 n_2'\bk}\mu_{n_1'\bk,n_2'\bk}^j A_{\lambda_1}^{n_1 n_1'} A_{\lambda_2}^{n_2 n_2'}\delta_{n_1 n_2}
\end{eqnarray}
While the above expression can be used to calculate the dipole matrix elements between two exciton states, it does not provide information on the optical selection rules for the relative-motion state of excitons. Instead, we use the exciton eigenstate representation in Eq.~(\ref{eq:exciton}) to obtain
\begin{eqnarray}
B^j_{\lambda_1 \lambda_2} & = & \frac{1}{N}\sum_{n_1 n_1' n_2 n_2'\bk\bk'}\left(\mu_{n_1\bk,n_2\bk}^j\delta_{n_1' n_2'}+\mu_{n_1'\bk',n_2'\bk'}^j\delta_{n_1 n_2}\right) \nn\\
& & \times\int d\mathbf{r}\phi_{\lambda_1}^{n_1n_1'}(r)\phi_{\lambda_2}^{n_2n_2'}(r)e^{i(\bk-\bk')\cdot\br}\nn\\
& & +\frac{e}{N}\sum_{n\bk n'\bk'}\int d\mathbf{r}\phi_{\lambda_1}^{nn'}(r)\br^j\phi_{\lambda_2}^{nn'}(r)e^{i(\bk-\bk')\cdot\br},
\end{eqnarray}
The three terms identify clearly the optical selection rules in terms of Bloch states of electrons, holes, or in terms of the excitonic relative-motion states. The reason is that, using the chain rule for $\bnabla_\bk$, the electric dipole moment operator can act either on the Bloch states of the electron or the hole, or on the excitonic relative-motion state.
Simplifying the above equation gives
\begin{eqnarray}
B^j_{\lambda_1 \lambda_2} & = & \frac{1}{N}\sum_{n_1 n_1' n_2 n_2'\bk}\left(\mu_{n_1\bk,n_2\bk}^j\delta_{n_1' n_2'}+\mu_{n_1'\bk,n_2'\bk}^j\delta_{n_1 n_2}\right) \nn\\
& & \times\int d\mathbf{r}\phi_{\lambda_1}^{n_1n_1'}(r)\phi_{\lambda_2}^{n_2n_2'}(r)\nn\\
& & +e\sum_{n n'}\int d\mathbf{r}\phi_{\lambda_1}^{nn'}(r)\br^j\phi_{\lambda_2}^{nn'}(r),
\end{eqnarray}
Here we can define the excitonic relative-motion dipole matrix element
\be
d_{n,n',\lambda_1,\lambda_2}^j=e\sum_{n n'}\int d\mathbf{r}\phi_{\lambda_1}^{nn'}(r)\br^j\phi_{\lambda_2}^{nn'}(r).
\ee
Note that the exciton transition matrix element $B^j_{\lambda_1 \lambda_2}$ contains an electron dipole matrix element, a hole dipole matrix element, and an excitonic relative-motion dipole matrix element. Each of these dipole matrix elements needs to satisfy optical selection rules in order to be nonzero. In 3D for isotropic systems, the hydrogen-like relative-motion wavefunctions follow the selection rule $\Delta l=\pm 1$ and $\Delta m_\pm=-1,0,+1$ in terms of orbital angular momentum $l$, and the two exciton states must have opposite parity \cite{Sakurai}. In particular, $\Delta m_\pm=\pm 1$ for right and left circularly polarized light, and $\Delta m_\pm=0$ for $z$-polarized light. The crystal field reduces the $SU(2)$ symmetry to the symmetry of that 3D crystal and the corresponding symmetries of the particular $k$-points.
In the case of anisotropic excitons in uniaxial crystals, the crystal field reduces the symmetry from $SU(2)$ to $D_{\infty h}$, and then to $D_{3h}$ at the $\Gamma$-point of TMCs. Therefore, the selection rules weaken to $\Delta m_\pm=\pm 1\pm 3\nu$ for right and left circularly polarized light, and $\Delta m_\pm=\pm 3\nu$ for $z$-polarized light, where $\nu$ is an integer \cite{Mak2018}.

Compared with the LR selection rules given by group theory, in the case of NLR we need to amend the LR selection rules by considering the symmetries of both the initial and final envelope wavefunctions for intermediate states. Then, group theory states that an electric dipole moment matrix element involving intermediate states is nonzero by symmetry only if the direct product of IRs $\Gamma(\left|n_1\bf{k}\right>)\otimes \Gamma(\left|n_2\bf{k}\right>)\otimes \Gamma(\left|\phi_{\lambda_1}\right>)\otimes \Gamma(\left|\phi_{\lambda_2}\right>)\otimes \Gamma(\mu^j)$ contains $\Gamma(I)$ in its decomposition in terms of a direct sum. Note that the initial and final exciton envelope states can be different. Therefore, the symmetries of the initial and final exciton envelope wavefunctions can change the optical selection rules imposed by symmetry when compared with the noninteracting electron-hole pairs.

In the case of anisotropic excitons, we find that the relative-motion wavefunctions follow the selection rule $\Delta m_\pm=-1,0,+1$ and the two exciton states must have opposite parity. In particular, $\Delta m_\pm=\pm 1$ for right and left circularly polarized light, and $\Delta m_\pm=0$ for $z$-polarized light. These optical selection rules are valid also in the 2D case.

Then the second-order susceptibility for excitons is given by
\be
\chi_{ijk}^{(2)}(\omega_\sigma,\omega_q,\omega_p) = \frac{N}{\epsilon_0}\mathcal{P}_F\sum_{\lambda_1,\lambda_2} \frac{B^i_{g\lambda_1} B^j_{\lambda_1\lambda_2} B^k_{\lambda_2g}f_{g\lambda_2\bk}}{(\hbar\omega_\sigma-E_{\lambda_1})(\hbar\omega_p-E_{\lambda_2})},
\label{eq:Chi2X}
\ee

In bulk GaSe, the energy difference between the VB and CB+1 is equal to 3.12 eV at the $\Gamma$-point in GW approximation in the IPA. The corresponding exciton binding energy is 0.87 eV, resulting in an exciton peak at 2.25 eV, corresponding to the $B_1^\parallel$ resonance peak shown in Fig.~\ref{fig:GaSe-bulk}.
Therefore, the photon energy must be below 2.25/2=1.125 eV to give rise to the off-resonant out-of-plane (z direction) optical transition driven by $\pi$-polarized photons ($\Delta m=0$).

\begin{widetext}
A similar derivation for the third-order NLR of excitons yields the excitonic susceptibility in third-order perturbation theory given by 
\be
\chi_{k j i h}^{(3)}\left(\omega_\sigma, \omega_r, \omega_q, \omega_p\right)=\frac{N}{\epsilon_0} \mathcal{P}_F \sum_{\lambda_1\lambda_2\lambda_3} \frac{B^k_{g\lambda_1} B^j_{\lambda_1\lambda_2} B^i_{\lambda_2\lambda_3} B^h_{\lambda_3g}f_{g\lambda_3\bk}}{\left(E_{\lambda_1}-\hbar\omega_\sigma\right)\left(E_{\lambda_2}-\hbar\omega_q-\hbar\omega_p\right)\left(E_{\lambda_3}-\hbar\omega_p\right)},
\label{eq:Chi3X}
\ee
where $\omega_\sigma=\omega_p+\omega_q+\omega_r$.
\end{widetext}
These formulas for the $\chi^{(2)}$ and $\chi^{(3)}$, derived by means of perturbation theory, provide an approximation to the non-perturbative results for $\chi^{(n)}$ obtained by means of the real-time KBE.

\section{Numerical Scheme}
\label{sec:numerics}
The numerical scheme for calculating LR and NLR non-perturbatively can be summarized by the following routine:

\begin{enumerate}

\item Calculate the charge  density, the KS eigenvalues and eigenfunctions from DFT.  

\item Calculate the QP corrections by means of the ev$GW$ approach.
 
\item Integrate the equation of motion based on the KBE \eqref{KBE} 
%
%
using KS wavefunctions and switching on the periodically oscillating electric field ${\mathbf{E}}\left( t \right)$ at $t=0$ in order to get the linear and nonlinear response of the system in $\mathbf{P}(t)$.

\item Calculate the excitonic polarization $\mathbf{P}(t)$ using Eq.~\eqref{P}.

\item Use Fast Fourier Transform (FFT) \cite{FFT} convert time-dependent $\mathbf{P}(t)$ into frequency-dependent $\mathbf{P}(\omega)$ in order to extract the linear $\chi^{(1)}$ and nonlinear susceptibilities $\chi^{(n)}$ from the total polarizability $\mathbf{P}(t)$ using Eq.~(\ref{eq:Polar}).   

\item Repeat (3) - (5) steps for varying frequency $\omega$ to obtain the linear $\chi^{(1)}$ and nonlinear susceptibilities $\chi^{(2)}$ and $\chi^{(3)}$, or, if desired, all nonlinear susceptibilities $\chi^{(n)}$, $n$ being a positive integer, within a given energy range.

\end{enumerate}

For DFT calculations we use the QUANTUM ESPRESSO (QE) ab initio simulation package \cite{QE1,QE2,QE}.
We use fully-relativistic optimized norm-conserving Vanderbilt (ONCV) pseudopotentials \cite{Hamann2013} from Schlipf-Gygi pseudopotential library \cite{Schlipf-Gygi2015,QS}. 
These pseudopotentials are optimized for $GW$ calculations with SOC \cite{Scherpelz2016} and are consistent with nonlinear calculations in YAMBO. 
The plane-wave energy cutoff was set to 450 eV. 
The energy and force convergence tolerance were set to $10^{-8}$ eV
and 0.001 eV/\AA, respectively.
The Brillouin zone of the structure is sampled by a $18 \times 18 \times 1$ $k$-mesh for ML and $18 \times 18 \times 4$ $k$-mesh for bulk which is enough to get qualitative LR and NLR results. 
For ML calculations, the layers are separated by 20 Å between layers to minimize image interaction between layers due to the periodic boundary condition. We also use a truncated Coulomb potential for ML calculations in order to achieve faster convergence and eliminate interaction between repeated layers. For ML calculations we use random the integration method implemented in YAMBO \cite{Sangalli2019} to avoid divergence in small $q$ in the Coulomb potential and obtain converged results for many-body calculations.     

DFT charge density and wave functions are used as a starting point for the $GW$ calculations. As often pointed out, the first step of $GW$, $G_0W_0$, shows inaccurate results for many systems. 
Another problem is that the $G_0W_0$ correction depends on the initial  charge density. 
This issue can be eliminated by utilizing the initial charge density obtained by using hybrid exchange-correlation functionals like PBE0 \cite{PBE0}, HSE \cite{HSE}, or range-separated hybrid functionals \cite{Wing2021,Zhan2023}. The latter show good results if they are used as starting point for $GW$ and BSE calculations \cite{Camarasa2023}.
In our approach, in order to avoid dependence on the initial charge density, we apply the self-consistent $GW$ method on eigenvalues only for both $G$ and $W$ (ev$GW$) implemented in the YAMBO package. 
In the self-consistent $GW$ calculations, it is necessary to update the wave functions as well. However, for many systems, the DFT wave functions are already quite accurate, and achieving self-consistency in terms of eigenvalues alone is often satisfactory \cite{Faber2013}. 
We found that for the studied systems three iterations of ev$GW$ is sufficient to reach convergence for band gap correction and eigenvalues within 5-10\% accuracy. Alternatively, if  the scissor correction is used for the first step $G_0W_0$, then the number of ev$GW$ iterations is reduced to two to reach the same accuracy. We also used the terminator technique developed by Bruneval and Gonze \cite{Bruneval2008} to achieve faster convergence with respect to sums over states. 

The many-body electron-hole interactions were calculated using the YAMBO many-body software \cite{Sangalli2019} with the included LUMEN package \cite{Attaccalite2011}. 
This approach has recently been successfully used to calculate the SHG in alloyed transition metal dichalcogenides, h-BN \cite{Lucking2018,Beach2020}, and in $\beta$-NP \cite{Kolos2021}.

\section{Results}
\label{sec:optical}

\subsection{Optical response in TMCs}

\begin{figure}
	\begin{center}
		\includegraphics[width=2.5in]{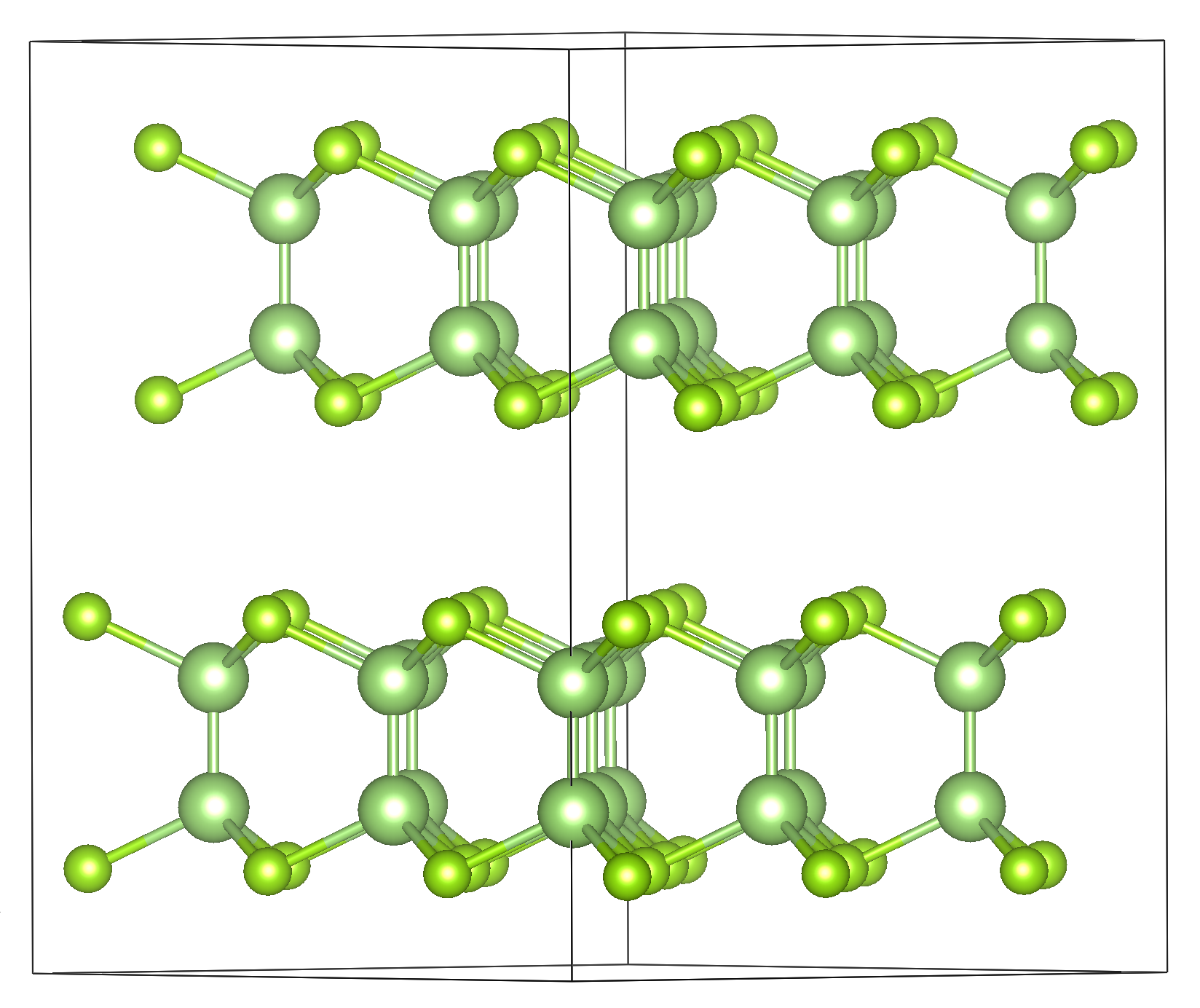}
	\end{center}
	\caption{AB stacking of layers in $\epsilon$-GaSe. Ga atoms are big green balls, and Se atoms are small.}
	\label{fig:GaSe_3D}
\end{figure}

\begin{figure*}
	\begin{center}
		\includegraphics{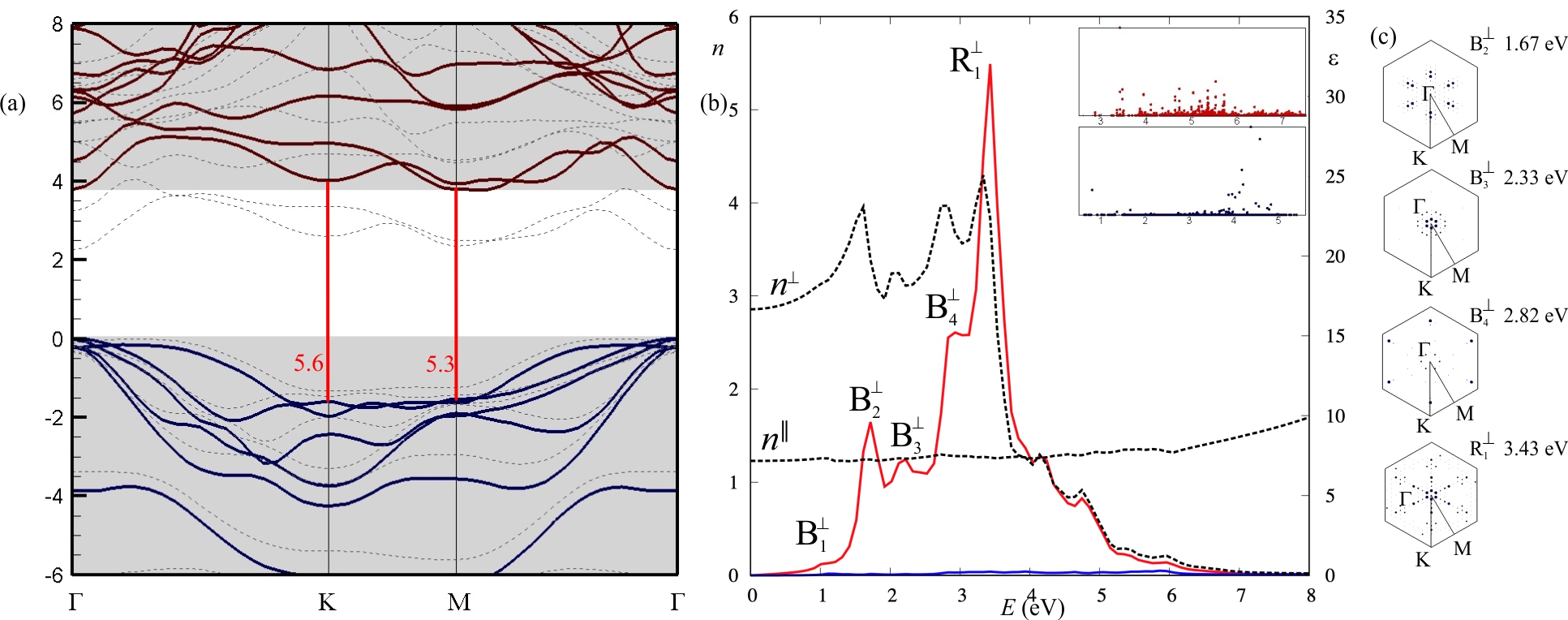}
	\end{center}
	\caption{GW-BSE results for ML GaSe: (a) GW-band structure (solid lines) and PBE-band structure (dashed lines) for comparison. (b) LR functions: refractive index n (dotted line) and imaginary part of the dielectric function Im$\epsilon^{(1)}$, red solid line for perturbation in $xy$ plane ($\perp$) and blue line for perturbation in $z$ direction ($\|$). (c) Excitonic peak energies and transition weights in k-space for B$_1^\parallel$, R$_1^\parallel$, R$_2^\parallel$, and R$_2^\perp$. The B$_1^\perp$ peak is at 1.04 eV with transition weight at $\Gamma$-point only.}
	\label{fig:GaSe-layer}
\end{figure*}


 \begin{figure*}
	\begin{center}
		\includegraphics{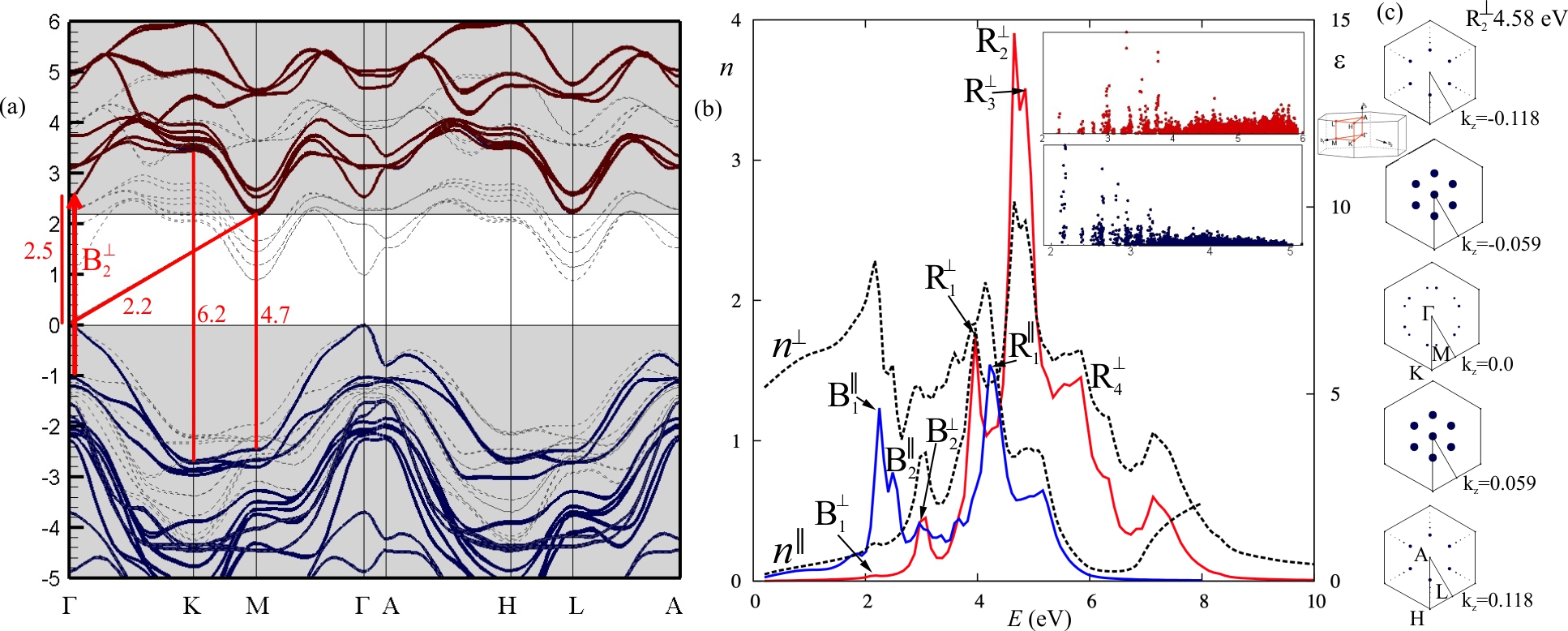}
	\end{center}
	\caption{GW-BSE results for bulk AB-stacked $\epsilon$-GaSe: (a) GW-band structure (solid lines) and PBE-band structure (dashed lines) for comparison.  (b) LR functions, the $B_2^\perp$ exciton peak is in excellent agreement with experiment. The insert shows dark-bright exciton statistics. (c) Selected excitonic peak energy and transition weights in k-space for R$_2^\perp$.}
	\label{fig:GaSe-bulk}
\end{figure*} 


 \begin{figure*}
	\begin{center}
		\includegraphics[width=6.5in]{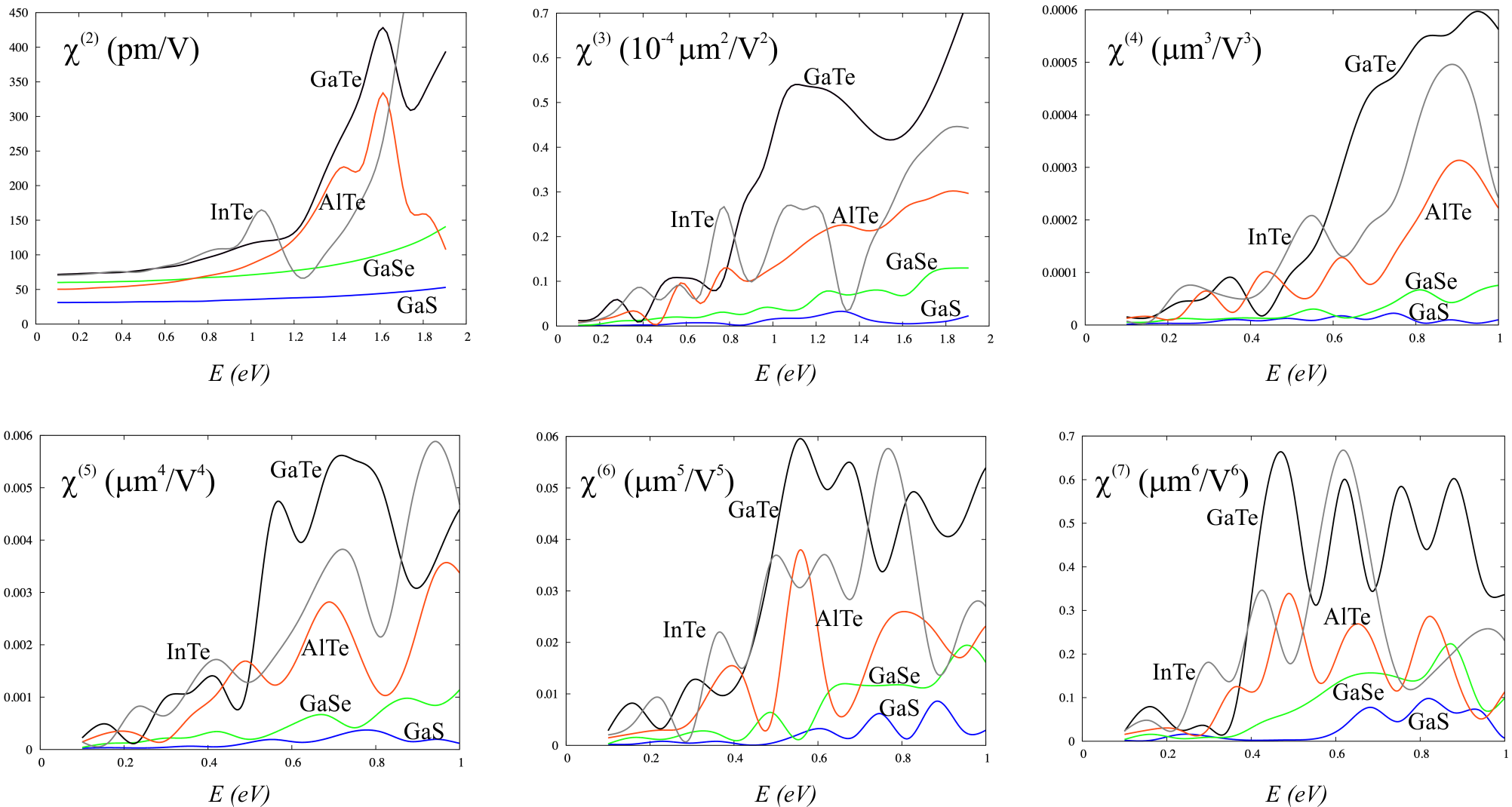}
	\end{center}
	\caption{Nonlinear coefficients $\chi^{(2)}$ - $\chi^{(7)}$ as a function of energy of the incident photon for TMCs.}
	\label{fig:X2X3}
\end{figure*}  

 \begin{figure}
	\begin{center}
		\includegraphics[width=3.4in]{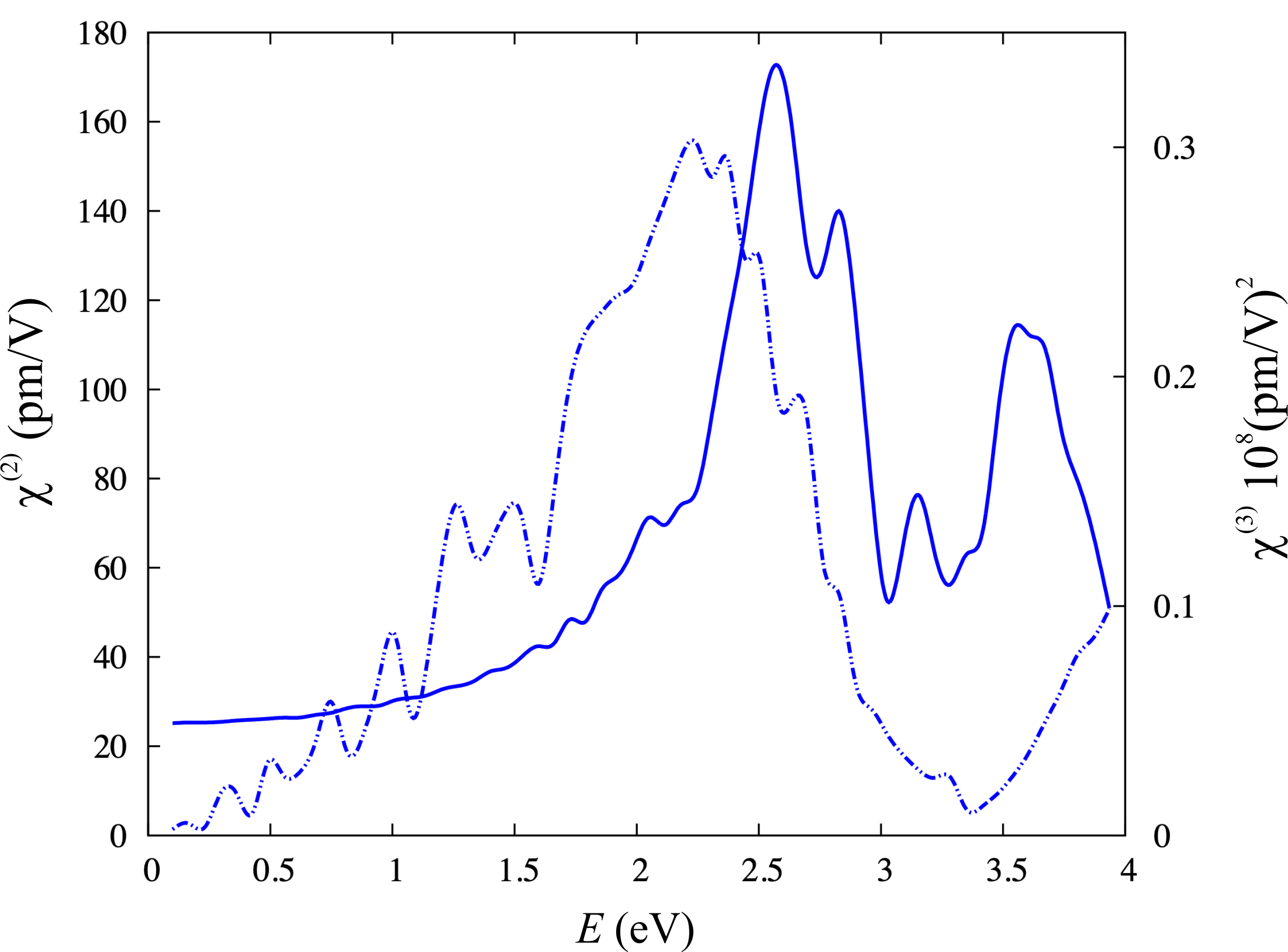}
	\end{center}
	\caption{Nonlinear coefficients $\chi^{(2)}$ (solid) and $\chi^{(3)}$ (dashed-dotted line) as a function of energy of the incident photon for $\epsilon$-GaSe.}
	\label{fig:X2X3_GaSe}
\end{figure}

\begin{figure*}
	\begin{center}
		\includegraphics{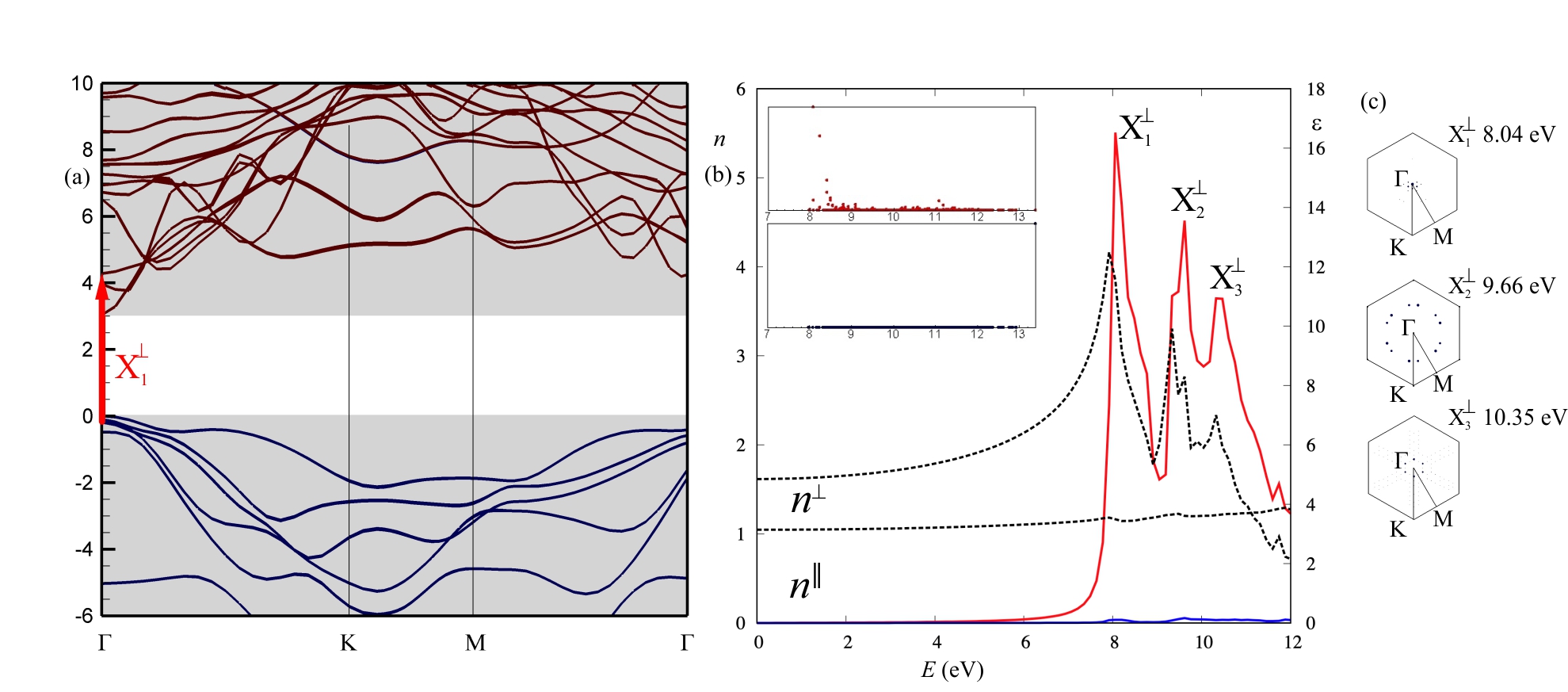}
	\end{center}
	\caption{GW-BSE results for GaS ML: (a) GW-band structure. (b) Imaginary part of the dielectric function, blue line $\|$ and red line $\perp$ component. Here we find anti-bound excitons X$_1^{\perp}$, X$_2^{\perp}$, and X$_3^{\perp}$, for which the repulsive exciton exchange interaction is stronger than the attractive exciton direct interaction, resulting effectively in a negative exciton binding energy. (c) Excitonic peak energies and transition weights in k-space for X$_1^{\perp}$, X$_2^{\perp}$, and X$_3^{\perp}$ peaks for GaS ML.}
	\label{fig:GaS}
\end{figure*} 


\begin{figure*}
	\begin{center}
		\includegraphics{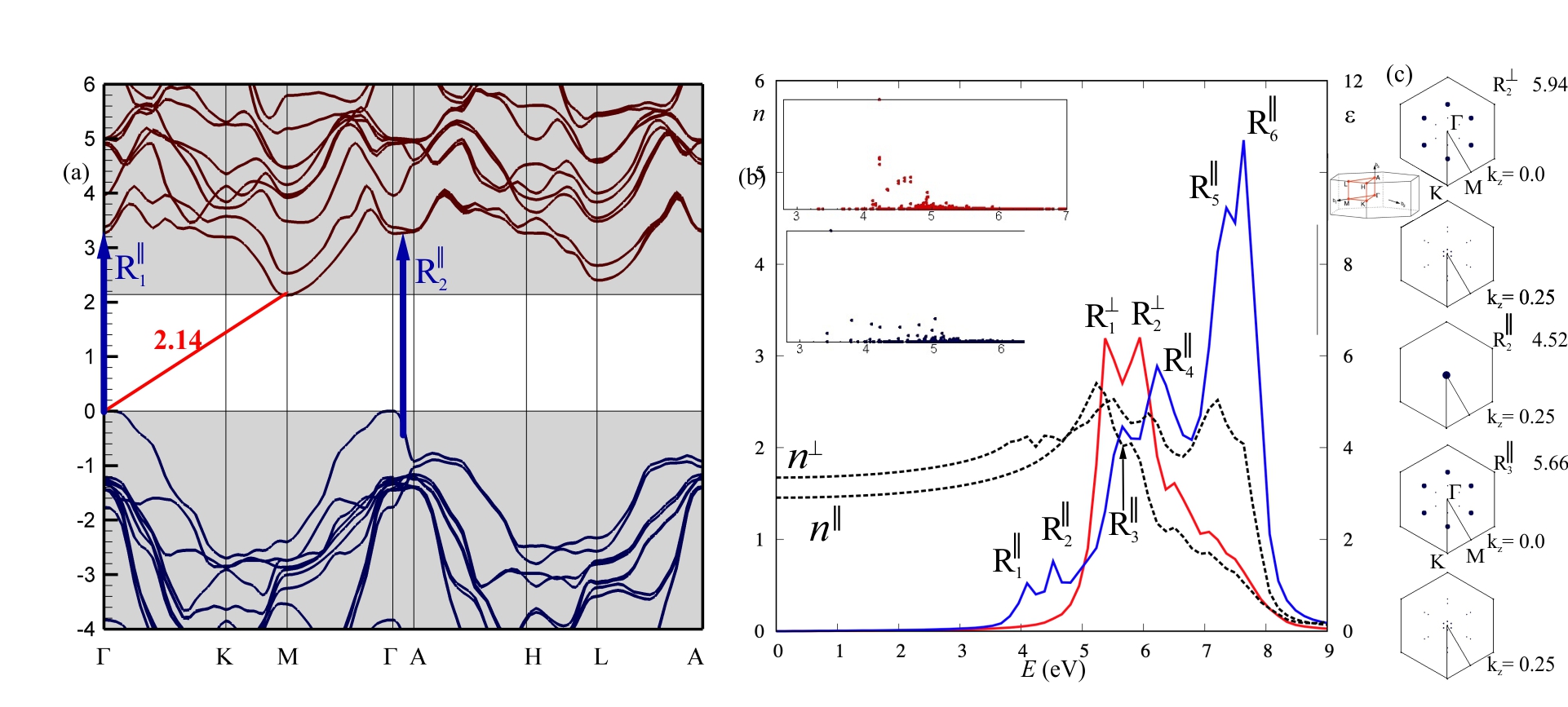}
	\end{center}
	\caption{(a) Band structure of bulk AB-stacked $\epsilon$-GaS. (b) Imaginary part of the dielectric function, blue line $\|$ and red line $\perp$ component. (c) Selected excitonic peak energies and transition weights in k-space.}
	\label{fig:e-GaS}
\end{figure*}

\begin{figure*}
	\begin{center}
		\includegraphics{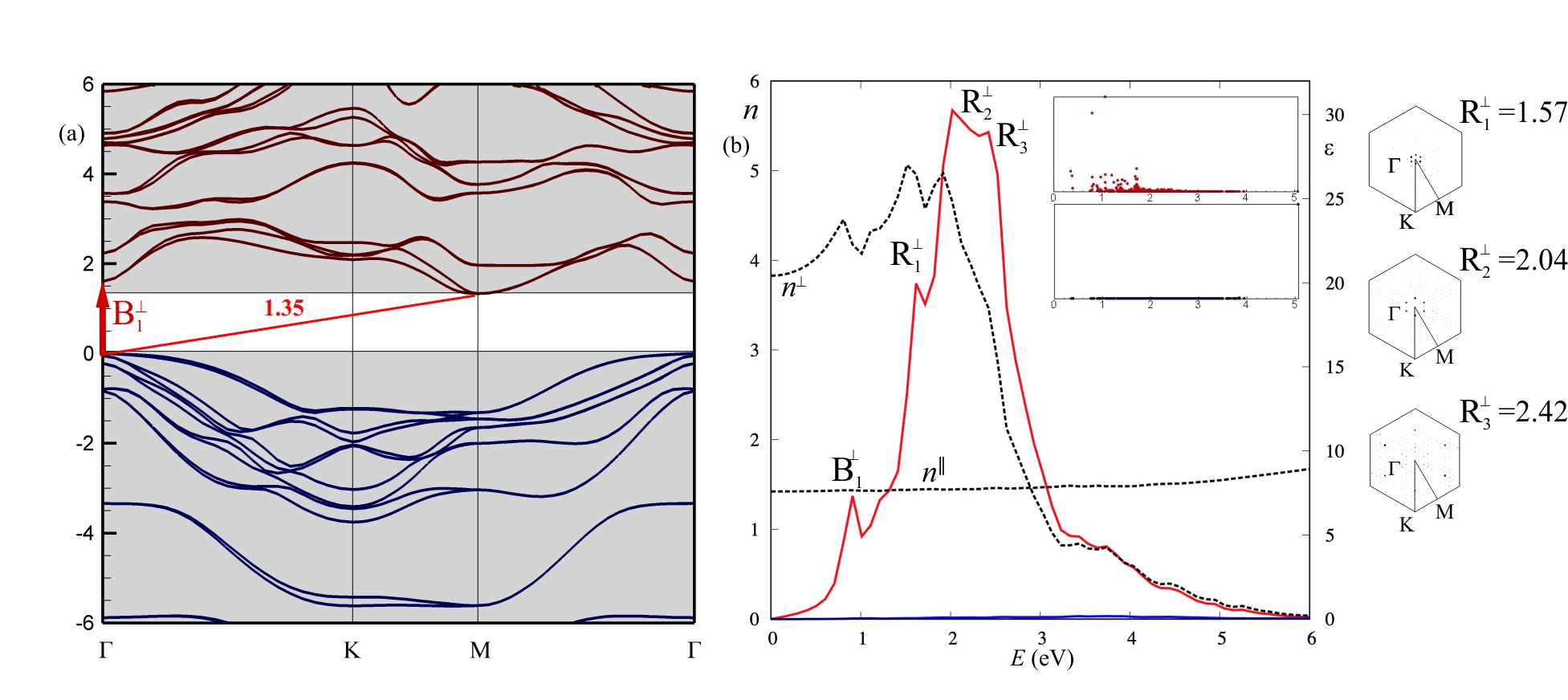}
	\end{center}
	\caption{(a) Band structure of GaTe ML. (b) Imaginary part of the dielectric function, blue line $\|$ and red line $\perp$ component.}
	\label{fig:GaTe}
\end{figure*} 

\begin{figure*}
	\begin{center}
		\includegraphics{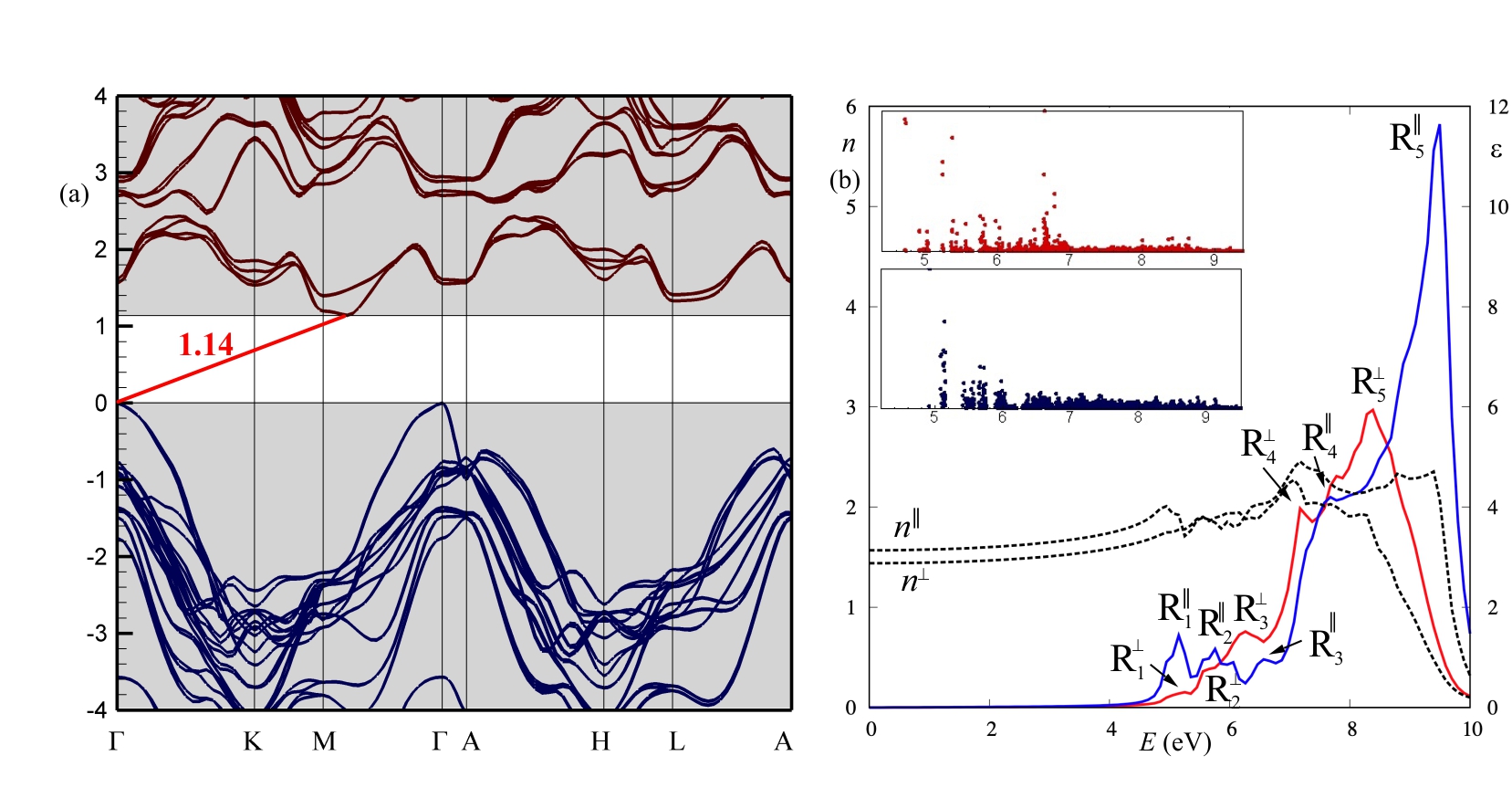}
	\end{center}
	\caption{(a) Band structure of bulk AB-stacked $\epsilon$-GaTe. (b) Imaginary part of the dielectric function, blue line $\|$ and red line $\perp$ component.}
	\label{fig:e-GaTe}
\end{figure*}

The crystal structure of $\epsilon$-GaSe crystal is shown in  Fig.~\ref{fig:GaSe_3D}. In AB stacking for bulk GaSe the distance between the layers is 7.96 \AA. Within the ML itself, the atoms are connected through covalent chemical bonds, whereas between the layers the connection is due to van der Waals forces.   
The bulk crystal in $\epsilon$ form has a band gap of 2.2 eV, and the band gap increases by decreasing the number of layers due to quantum confinement. For one ML the band gap is 3.8 eV. 
By substituting Se atoms by heavier elements from the same column in the Periodic Table, the band gap decreases. Thus, for GaTe ML, the band gap is 1.65 eV, and for InTe the band gap is 0.6 eV. 
The $GW$ corrected band gaps of studied materials for ML and for bulk with AB stacking ($\epsilon$ form) are presented in Table \ref{table:1}. 

The bandstructure and LR for ML and bulk GaSe are presented in Fig.~\ref{fig:GaSe-layer} and Fig.~\ref{fig:GaSe-bulk}, correspondingly. The bandstructure of ML GaSe exhibits an indirect band gap at the $\Gamma$ point of 3.8 eV, whereas the bandstructure of bulk GaSe exhibits an indirect band gap between the $\Gamma$-M of 2.2 eV. The experimental measurement by angle-resolved photoemission spectroscopy of band gap for ML shows a band gap of 3.5 eV \cite{BenAziza2017}, whereas the experimental measurement of the out-of-plane response for the $\pi$ transition shows an energy gap of 2.13 eV \cite{Toullec1980}. Black dashed lines on Fig.~\ref{fig:GaSe-layer} and Fig.~\ref{fig:GaSe-bulk} denote DFT and solid lines denote ev$GW$ results. 
For LR functions the refractive index $n$ (dotted line) and the imaginary part of the dielectric function Im[$\epsilon^{(1)}$] (blue and red solid lines) are plotted for two electric field excitations, in plane ($\perp$) and out of plane ($\|$). 

The LR spectra are calculated by solving the BSE using YAMBO. The exciton peaks $B_1$ and $B_2$ correspond to bound exciton states, whereas the $R$ peaks correspond to resonant exciton states with total energy above the band gap and positive exciton binding energy, i.e. they are bound excitons. The bound state $B_2^{\perp}$ corresponds to the VB-1 $\rightarrow$ CB transition at the $\Gamma$ point marked by the red arrow in Fig.~\ref{fig:GaSe-bulk}a. The calculated binding energy of the $B_2^{\perp}$ exciton is 0.32 eV, which is in excellent agreement with the experimental measurement of the exciton at 3.378 eV excited with in-plane polarized light and with an exciton binding energy of 318 meV  \cite{Zalamai2020}.     

\begin{center}
\begin{table}
\begin{tabular}{ c c c  }
\hline
            Compound & ML  & Bulk  \\
\hline
            GaS  & 2.60  & 2.14     \\
            GaSe & 3.80  & 2.20 \\
            GaTe & 1.35  & 1.14 \\
\hline
        \end{tabular}
    \caption{$GW$ corrected band gap in eV. }
    \label {table:1}
    \end{table}
    \end{center}

\begin{center}
\begin{table}
\begin{tabular}{ c c   }
\hline
            Compound & $\chi^{2} (pm/V)$     \\
\hline
            GaS  & 34.2    \\
            GaSe  & 68.1    \\
            GaTe  & 101.6    \\
            AlTe  & 73.2    \\
            InTe  & 108.8   \\
\hline
        \end{tabular}
    \caption{Nonlinear coefficients $\chi^{(2)}$ at $E=0.86$ eV. }
    \label {table:2}
    \end{table}
    \end{center}

\begin{center}
\begin{table}
\begin{tabular}{ c c c c c    }
\hline
            Peak & Energy & Transition & $E_X$ & k-point   \\
\hline
ML GaSe \\
\hline
$B_1^{\perp}$ & 1.04 & $VB\rightarrow CB$ & 2.76  &  $\Gamma$   \\
$B_2^{\perp}$ & 1.67 & $VB\rightarrow CB$ & 3.33  &     \\
$B_3^{\perp}$ & 2.33 & $VB-1\rightarrow CB$ & 0.74  &     \\
$B_4^{\perp}$ & 2.82 & $VB\rightarrow CB+1$ & 0.74  &     \\
$R_1^{\perp}$ & 3.43 & $VB-2\rightarrow CB+1$  & 0.69  &     \\
\hline
bulk GaSe \\
\hline
$B_2^{\perp}$ & 3.28 & $VB-1\rightarrow CB$ & 0.32 & $\Gamma$    \\
$B_1^{\|}$ & 2.25 & $VB\rightarrow CB+1$ & 0.87 & $\Gamma$    \\
$B_2^{\|}$ & 2.49 & $VB\rightarrow CB$ & 0.08 & $\Gamma$    \\
\hline
ML GaS \\
\hline
$X_1^{\perp}$ & 8.04 & $VB-1\rightarrow CB+2$ & -3.79 &     \\
$X_2^{\perp}$ & 9.66 & $VB-1\rightarrow CB$ & -2.46  &     \\
$X_3^{\perp}$ & 10.35 & $VB\rightarrow CB+1$ & -6.45  &     \\
\hline
ML GaTe \\
\hline
$B_1^{\perp}$ & 0.88 & $VB\rightarrow CB$ & 0.72 & $\Gamma$    \\
$R_1^{\perp}$ & 1.57 & $VB-2\rightarrow CB$ & 0.70 &    \\
$R_2^{\perp}$ & 2.04 & $VB-1\rightarrow CB$ & 0.68 &   \\
$R_3^{\perp}$ & 2.42 & $VB-1\rightarrow CB$ & 0.72 &   \\  
\hline
\hline
\end{tabular}
\caption{Peak, energy of the peak, transition, exciton binding energy (eV), and k-point corresponding to transition.}
\label {table:T3}
\end{table}
\end{center}


The NLR spectra are calculated by solving the dynamical real-time KBEs using YAMBO. The NLR parameters for the studied materials are presented in Fig.~\ref{fig:X2X3}. For GaS the nonlinear coefficients $\chi^{(2)}$ in the range of 0-2 eV is around 30 pm/V (see Fig.~\ref{fig:X2X3}). 
Substituting the Ga atom by the lighter Al atom decreases the NLR (see AlTe curve in Fig.~\ref{fig:X2X3}), whereas for InTe the NLR is increased in the off-resonant region, i.e. lower than band gap; however, the band gap for this compound is 0.6 eV. Table \ref{table:2} shows nonlinear coefficients $\chi^{(2)}$ corresponding to the energy of incident light of $E=0.86$ eV. The $\chi^{(2)}$ coefficient is increased for the material sequence Gas-GaSe-GaTe from 34.2 to 101.6 pm/V. In the case of the material sequence AlTe-GaTe-InTe $\chi^{(2)}$ is increased from 73.2 to 108.8 pm/V. 

The high-harmonic nonlinear coefficients $\chi^{(3)}$ - $\chi^{(7)}$ for the studied materials are also presented in Fig.~\ref{fig:X2X3}. The $\chi^{(n)}, n=3...7$ coefficients oscillate with respect to the energy of the photon corresponding to possible  multiple many-photon transitions. In general, we find the same tendency, i.e. when substituting lighter elements by heavier ones, the NLR $\chi^{(n)}$ is increased as well. The efficiency of nonlinear signal depends on the intensity of incident photons according to Eq.~(\ref{eq:Polar}) and can be estimated for $\chi^{(3)}$ and $\chi^{(4)}$ as $10^{-6}$ with laser intensity of $10^{6}$ W/cm$^2$, whereas for a laser intensity of $10^{8}$ W/cm$^2$, the efficiency of the nonlinear signal is estimated to be around $10^{-2}$.  

The band structures and LR of ML and bulk GaS are presented in Fig.~\ref{fig:GaS} and Fig.~\ref{fig:e-GaS}, respectively. Experiments show a direct band gap of bulk GaS of 3.04 eV, whereas the indirect band gap is 2.53 eV \cite{Aulich1969,Ho2006}. These results are in good agreement with our $GW$ calculation resulting in 3.12 eV and 2.14 eV, respectively. The experimental measurement of $\chi^{(2)}$ for GaS at 1.4 eV excitation shows a value of 48 pm/V \cite{Ahmed2022}, which is in very good agreement with our calculated value of 40 pm/V (see Fig.~\ref{fig:X2X3}).   Fig.~\ref{fig:e-GaTe} and Fig.~\ref{fig:GaTe} show band structures and LR of ML and bulk GaTe, respectively. Exciton binding energy (in eV) and transitions for ML GaTe corresponding to peaks at Fig.~\ref{fig:GaTe}b are shown in Table \ref{table:T3}. The direct band gap of ML GaTe at the $\Gamma$-point is 1.60 eV, whereas the indirect band gap in the $\Gamma$-M region is 1.35 eV. These results are in very good agreement with experimental measurements for the direct band gap of 1.67 eV \cite{Singha2022}. Susoma et.al. \cite{Susoma2016} provide experimental measurement of THG, giving $\chi^{(3)}=2 \times 10^{-16}$ (m/V)$^2$ for GaTe at 2.45 eV. Our calculated value for $\chi^{(3)}$ is $0.7 \times 10^{-16}$ (m/V)$^2$ at 1.9 eV (see Fig.~\ref{fig:X2X3}), which is in good agreement with their experiment.

\subsection{Discussion}

\subsubsection{Screening Effects and Exciton Nature in 2D and 3D TMCs}

In semiconductor materials, excitons---bound states of an electron and a hole---play a crucial role in determining the optical properties of the material. Excitons can be categorized into two types: bright excitons and dark excitons, which differ primarily in their ability to interact with light. The optical selection rules for the Bloch states and for the exciton envelope wavefunctions provide the main conditions for the distinction between bright and dark excitons. If the optical transition is allowed, the exciton is bright. If the optical transition is forbidden, then the exciton is dark. A second important condition is due to the fact that a photon's momentum is very small compared to the possible momentum difference between the Bloch states of an electron and a hole in the first Brillouin zone, which is also the reason why the electric dipole approximation is often applied. In this approximation, only direct excitons are bright, whereas indirect excitons are dark, unless exciton-phonon coupling is considered, which is beyond the scope of this paper. 

Analyzing the BSE data for TMCs, we observe a significant difference in the nature of the lowest-energy excitons between 2D ML TMCs and their 3D bulk counterparts. Specifically, the lowest-energy excitons in 2D ML TMCs tend to be dark, whereas in 3D bulk TMCs, these excitons are often bright or exhibit a mixed bright-dark character. This behavior is primarily attributed to the differences in the screening environment between 2D and 3D systems and how screening affects the exciton orbital angular momentum.

In 2D ML TMCs, the lowest-energy excitons are typically dark. These dark excitons are characterized by high orbital angular momentum. Due to the optical selection rules and the large angular momentum, they do not couple efficiently with light, making them optically inactive (see Sec.~\ref{sec:selection_rules_excitons}).

One of the key factors behind this behavior is the unique nature of distance-dependent screening in 2D materials. In 2D systems, the Coulomb interaction between an electron and a hole is screened less effectively than in 3D bulk materials, in particular this screening depends on the in-plane distance between the electron and hole. The reduced screening enhances the strength of the Coulomb interaction at large distance, which has two major effects: it increases the binding energy of the exciton, and it favors excitonic states with large orbital angular momentum.

In 2D systems, excitons with large orbital angular momentum are less prone to radiative recombination because the recombination process requires a change in orbital angular momentum that is suppressed in such states. This makes these high-orbital angular-momentum excitons optically inactive, i.e. dark. As a result, dark excitons with large orbital angular momentum often constitute the ground state in ML TMCs.

In contrast, in 3D bulk TMCs, the lowest-energy excitons are typically bright or exhibit a mixed bright-dark character. 
The key difference in 3D materials is the nature of the dielectric screening. In bulk systems, screening is uniform and more effective compared to 2D systems, reducing the overall Coulomb interaction between the electron and hole. This uniform, constant screening environment results typically in s-wave excitons with zero orbital angular momentum.

This trend has important implications for the optoelectronic properties of TMCs. In 2D TMCs, where dark excitons with large orbital angular momentum dominate, non-radiative recombination pathways may limit the efficiency of light emission, affecting applications such as light-emitting diodes (LEDs) and lasers. However, dark excitons can be useful in solar cells, where prevention of recombination of excitons is sought. Conversely, the prevalence of bright excitons in 3D bulk TMCs makes them more suitable for optoelectronic applications that rely on efficient photon emission, such as LEDs.

\subsubsection{Excitons in ML GaS: Dominance of the Exchange Interaction}

Typically, excitons are electron-hole pairs bound by Coulomb interactions, and their properties are determined by two primary contributions: the attractive direct interaction between the electron and hole, and the repulsive exchange interaction. In most materials, the balance between these interactions defines the exciton binding energy, which is typically positive, indicating a bound state.

In monolayer (ML) GaS, however, our calculated BSE data shown in Fig.~\ref{fig:GaS} and Table~\ref{table:T3} reveals a new and surprising result. For three excitons, $X_1^\perp$, $X_2^\perp$, and $X_3^\perp$, the exciton binding energy is found to be negative, suggesting that the repulsive exchange interaction dominates over the attractive direct interaction. This is an unusual scenario, as it implies that the exciton state is destabilized by the exchange interaction to the extent that it overcomes the binding effect of the direct Coulomb attraction.

This result highlights the importance of the exchange interaction in determining excitonic properties in two-dimensional materials. It suggests that, under certain conditions, excitons in materials like ML GaS can exhibit unconventional behavior, where the repulsive forces play a more prominent role than previously anticipated.

The dominance of the exchange interaction in ML GaS has several implications. First, it challenges the traditional understanding of exciton binding in 2D semiconductors. Second, it suggests that careful consideration of both direct and exchange interactions is necessary for accurately predicting excitonic behavior, especially in materials with complex electronic structures. 

\subsubsection{Nonlinear Response in 3D and 2D TMCs}

Our results showing the increase of $\chi^{(n)}$ when replacing lighter by heavier atoms can be understood in terms of the values of the electric dipole moments for $n=1$ by means of the formula in Eq.~(\ref{eq:Chi1X}), and also in terms of the detuning energies for $n>1$ by means of the formulas in Eqs.~(\ref{eq:chi2}) and (\ref{eq:chi3}) in IPA. For $n=2,3$  we provide a generalization of the NLR in second- and third-order nonlinear perturbation theory for exciton states in Eqs.~(\ref{eq:Chi2X}) and (\ref{eq:Chi3X}), which can be generalized to $n>3$ in a straightforward way. The formulas for the NLR in IPA agree with the formulas shown in Refs.~\cite{Ghahramani1991,Sipe1993,Aversa1995,Sharma2004}. 

Using this perturbative analysis, one notices that the linear and nonlinear coefficients are directly proportional to the value of the electric dipole moments of the semiconductor material. When substituting lighter elements by heavier ones, the electric dipole moment, i.e. polarization of the system, is increased due to the larger polarizabilities of electrons in the outer shells of the heavier atoms. Consequently, the larger electric dipole moments increase not only the LR, but also the NLR for $\chi^{(n)}$. 

For NLR the detuning of the photon's energy with respect to the resonant energies in the semiconductor material also matters. Since the nonlinear coefficients are inversely proportional the detuning energies, they increase with decreasing detuning energy. 

When substituting lighter elements by heavier ones, the band gap of the system is decreased, and consequently, the reduced band gap results in the increase of the coefficients $\chi^{(n)}$. 
The decrease in band gap in semiconductor materials when lighter atoms are replaced by heavier atoms within the same column of the periodic table, such as in the sequence GaS -> GaSe -> GaTe, can be explained by considering the following factors:
\begin{itemize}
\item Atomic Size and Bond Length: Heavier atoms generally have larger atomic radii compared to lighter atoms in the same group. This increase in atomic size leads to longer bond lengths in the crystal lattice. Longer bond lengths typically result in weaker bonding between atoms, which affects the energy levels of the valence and conduction bands.
\item Electronegativity: Heavier atoms, such as Se and Te, are generally less electronegative compared to lighter atoms like S. Lower electronegativity results in a reduction in the energy difference between the bonding (valence band) and antibonding (conduction band) states. This reduction contributes to a narrower band gap.
\item Spin-Orbit Coupling: Spin-orbit coupling becomes more significant with increasing atomic number. Heavier atoms like Te have stronger spin-orbit interactions compared to lighter atoms like S. This stronger spin-orbit coupling can lower the energy of the conduction band minimum, effectively reducing the band gap.
\item Orbital Overlap and Hybridization: The orbitals of heavier atoms (such as the p orbitals of Se and Te) are more spread compared to those of lighter atoms (such as S). This affects the overlap of atomic orbitals, leading to changes in the bonding character and the electronic structure of the material. Weaker orbital overlap in heavier atom compounds typically leads to a smaller band gap.
\end{itemize}

In general, we observe the overall trend that the MLs have larger band gaps than their corresponding bulk materials, which is due to confinement.

In summary, both the increase in the electric dipole matrix elements and the decrease of the band gap by substitution of lighter by heavier elements in the semiconductor alloy result in the increase of the LR and NLR functions $\chi^{(n)}$ (see Fig.~\ref{fig:X2X3}). Many authors notice large deviations in their results when measuring experimentally nonlinear response, sometimes even few orders of magnitude \cite{Saynatjoki2017,Fu2024}. This experimental broadening of the results makes it challenging to see the trend and rules in the nonlinear response. Nevertheless, Bredillet et.al. \cite{Bredillet2020} provide high accuracy measurement of nonlinear response from two-dimensional MoS$_2$ and WS$_2$ and measure $\chi^{(2)}$ coefficients of 130 and 530 pm/V, respectively. This experiment confirms our theoretical findings, namely that by substituting the lighter Mo element by the heavier W from the same column of the periodic table, the nonlinear response is increased. 

\subsubsection{Applications of High-Harmonic Generation}
GaSe crystals have already been used with an input pulsed laser at a center wavelength of 2.1 $\mu$m to demonstrate
High-Harmonic Generation (HHG) up to 10th order, i.e. 210 nm, in the deep-UV range well above the band gap of GaSe \cite{Imasaka2022}. Their results hold great promise for attosecond pulse
generation through solid-state HHG. A recent review on HHG and attosecond science can be found in Ref.~\cite{Li2020}. HHG is a nonlinear optical process that has emerged as a versatile tool with applications in various fields, ranging from quantum information science to materials characterization and laser technologies. Below, we highlight some of the key applications of HHG, particularly in quantum transduction, lasers, spectroscopy, and imaging.

HHG is a promising technique for quantum transduction, enabling the conversion of quantum information between photons with vastly different frequencies. This is crucial for bridging the frequency gap between microwave and optical photons in quantum networks \cite{Kimble2008, Wehner2018}. HHG can act as a frequency bridging tool by generating photons at harmonics of the original frequencies, potentially facilitating quantum communication over long distances \cite{Mirhosseini2020}.

While it is still challenging to reach the X-ray regime by means of solid-state HHG, it is worth mentioning that HHG is a key technology in the development of ultrafast lasers and optical frequency combs:
HHG enables the generation of coherent light in the extreme ultraviolet (XUV) and soft X-ray regions, producing ultrafast laser pulses in the attosecond regime. These attosecond pulses are critical for studying ultrafast electron dynamics in atoms and molecules \cite{Krausz2009, Nisoli2017}, with important applications in materials science and chemistry.
HHG can be used to generate frequency combs across a wide range of frequencies, making it a powerful tool for high-precision spectroscopy, timekeeping, and applications such as GPS systems and atomic clocks \cite{Udem2002, Cundiff2003,Ludlow2015}. The ability to generate high harmonics enhances the flexibility of frequency combs for diverse applications.

HHG is one of the most efficient methods for generating coherent light in the XUV and soft X-ray regions. These sources have several applications. For example, HHG provides a compact, tabletop alternative to large-scale X-ray Free-Electron Lasers (XFELs) \cite{Pellegrini2016}, offering a cost-effective solution for generating high-frequency coherent light. These sources are useful for materials characterization, molecular imaging, and probing electronic structures at atomic resolution.
Attosecond spectroscopy is based on high-harmonic generated attosecond pulses, which are essential for capturing ultrafast processes, such as electron dynamics within atoms and molecules \cite{Calegari2016, Nisoli2017}. This enables a deeper understanding of quantum phenomena on extremely short timescales.

HHG plays a crucial role in high-resolution imaging, particularly in the XUV and X-ray regimes.
For example, X-ray microscopy is enabled by their short wavelengths that allow for nanoscale resolution imaging, which is valuable in fields such as nanotechnology, biology, and materials science \cite{Cao2024}.
Another example is Coherent Diffractive Imaging (CDI), which is used to reconstruct high-resolution images from diffraction patterns \cite{Miao2015}. This technique is particularly important for imaging complex biological structures and nanomaterials with atomic precision.

HHG enables new forms of spectroscopy and material analysis, particularly in the XUV and soft X-ray regimes.
Ultrafast spectroscopy is based on high-frequency harmonics that allow for ultrafast probing of electronic transitions in materials, providing insights into electron dynamics, photoionization, and electron-lattice interactions \cite{Popmintchev2012}. HHG-generated XUV and X-ray photons can be used for element-specific spectroscopy, enabling the selective study of specific atoms or elements in complex materials and chemical reactions \cite{Sension2020}.

\section{Conclusion}
In this paper we consider first-principles methods to describe the linear and the nonlinear responses in 2D layered semiconductor materials. The approaches under consideration are the Bethe-Salpeter equation (BSE), the $GW$ approximation, the COHSEX approximation, and the Kadanoff-Baym equations (KBEs). While the BSE can be used to calculate the bound and resonant exciton states in equilibrium for the purpose of extracting the LR due to excitons, we use the KBEs to find the non-equilibrium NLR due to excitons including many-body interactions in a semiconductor material. Our theoretical results and predictions allow to develop several rules and strategies to build semiconductor materials with larger nonlinear response.    
We have focused on the NLR for the family of 2D layered structures made of IIA-VIA group TMC compounds, like GaS, GaSe, GaTe, InTe.
We show that by substituting lighter atoms with heavier atoms from the same column of the periodic table, i.e. 
S-Se-Te, and by reducing the detuning energy the NLR of the TMC crystals is increased, not only for $\chi^{(2)}$ and $\chi^{(3)}$, responsible for SHG and THG, but also for $\chi^{(n)}$ nonlinear response with $n>3$, giving rise to high harmonic generation in 2D semiconductor materials. Moreover, for example, when substituting Ga atoms in GaTe by lighter Al atoms, NLR is decreased in AlTe, whereas, substituting by heavier In atoms, NLR is increased in InTe. The obtained results can be explained by empirical models based on electric dipole matrix elements and the band gaps of the semiconductor materials.    
For that, we developed a detailed analysis of the linear and nonlinear optical selection rules by means of group and representation theory, showing strong connection to crystal symmetry and orbital characters of the bands and providing a method to predict the strength of linear and nonlinear responses of new materials. In particular, we derived general formulas for the nonlinear optical response based on exciton states in semiconductor materials. 
We identified bright and dark excitons. A detailed analysis of the dark excitons will be published elsewhere. The calculated binding energy of the $B_2^{\perp}$ exciton in GaSe is 0.32 eV, which is in excellent agreement with the experimental measurement of the exciton at 3.378 eV excited with in-plane polarized light and with an exciton binding energy of 318 meV  \cite{Zalamai2020}. Moreover, we find anti-bound excitons in ML GaS, which we attribute to dominant exciton exchange interaction.
We achieve good to excellent agreement with experimental measurements of linear exciton spectra and nonlinear coefficients for ML and bulk GaS, GaSe, and GaTe for $\chi^{(n)}$, $n=2$ and $n=3$. We predict the nonlinear response of InTe and AlTe for $2<n<7$.
The HHG regime has so far been rarely considered, both theoretically and experimentally. We predict high values of SHG, THG, and HHG for the TMC family of materials for $2\le n\le 7$, which is in agreement with similar trends in TMDs.  

\begin{center}
\begin{table*}
\begin{tabular}{ c c c c c c c  }
\hline
            Compound & $\perp$ & transition & degeneracy & $\|$ & transition & degeneracy \\
\hline
ML GaSe &  (1.04) & $VB\rightarrow CB$ & 3 & 1.04 & $VB\rightarrow CB$ & 3 \\
        &  1.15 & $VB\rightarrow CB$ & 1 & (1.15) & $VB\rightarrow CB$ & 1 \\
\hline
bulk GaSe & (2.25) & $VB\rightarrow CB$ & 2 & (2.11) & $VB-1\rightarrow CB$  & 1\\
          & 2.25   & $VB\rightarrow CB$ & 2 &  2.14  & $VB-1\rightarrow CB$  & 2 \\
\hline
ML GaS &  7.79  & $VB\rightarrow CB+2$ & 4 & 7.79 & $VB\rightarrow CB+2$ & 4 \\
       & (7.85) & $VB-1\rightarrow CB+2$ & 4 & (7.96) & $VB\rightarrow CB+2$ & 1 \\
\hline
ML GaTe & 0.88 & $VB\rightarrow CB$ & 4 & 0.88 & $VB\rightarrow CB$  & 4 \\
        & (0.89) & $VB-1\rightarrow CB$ & 1 & (0.93) & $VB-1\rightarrow CB$  & 1   \\
\hline
\end{tabular}
\caption{The lowest energy dark excitons (eV) and the corresponding transition and degeneracy. The lowest energy bright exciton is in parenthesis. }
\label {table:T4}
\end{table*}
\end{center}

\section{Acknowledgments}
M. N. L. acknowledges support by the Air Force Office of Scientific Research (AFOSR) under award no. FA9550-23-1-0455.
M. N. L. and D. R. E. acknowledge support by the AFOSR under award no. FA9550-23-1-0472 and DARPA under grant no. HR00112220011.
Preliminary calculations were performed at the Stokes high performance computer cluster of the University of Central Florida.
The final calculations were performed on the Pittsburgh high performance computer cluster provided by the ACCESS program of the National Science Foundation (NSF).
M. N. L. and D. S. acknowledge support by the NSF ACCESS program under allocation no. PHY230182.

\bibliographystyle{apsrev4-1}

\end{document}